\documentstyle[mnextra]{mn}
\input{epsf}
\def\twiddle{$\tilde{\hphantom{g}}$}
\def\WWW{http://star-www.dur.ac.uk/\twiddle cole/mocks/main.html }
\def\deg{\circ}
\def\hr{{\rm hr}}
\def\min{{\rm m}}

\def\BJ{B_{\rm J}}
\def\zmax{z_{\rm max}}
\def\zrest{z_{\rm rest}}
\def\zcut{z_{\rm cut}}
\def\zgal{z_{\rm gal}}
\def\r{{\bf r}}

\def\k{{\bf k}}

\def\etal{et al.\ }

\def\nbody{$N$-body\ }
\def\lsim{\mathrel{\hbox{\rlap{\hbox{\lower4pt\hbox{$\sim$}}}\hbox{$<$}}}}
\def\gsim{\mathrel{\hbox{\rlap{\hbox{\lower4pt\hbox{$\sim$}}}\hbox{$>$}}}}
\def\hmpc{h^{-1}\,{\rm Mpc}}
\def\Mpc{h^{-1}\,{\rm Mpc}}

\def\hkpc{h^{-1}\,{\rm kpc}}

\def\kms{{\rm km\,s^{-1}}}
\def\eg{{\rm e.g.\ }}
\def\ie{{\rm i.e.\ }}
\def\Omegab{\Omega_{\rm b}}
\def\COBE{{\it{COBE }}}
\def\sigmag{\sigma^{\rm gal}}
\def\sigmam{\sigma^{\rm mass}}

\nobrackets    
\seceqnum      
\begin{document}

\title[Mock Redshift Surveys]{Mock 2dF and SDSS galaxy redshift surveys}
\author[S. Cole \etal]{Shaun Cole$^{1,3}$, Steve Hatton$^{1,4}$, David H. Weinberg$^{2,5}$ and Carlos S. Frenk$^{1,6}$ \\
$^1$Department of Physics, University of Durham, Science
Laboratories, South Rd, Durham DH1 3LE. \\
$^2$Department of Astronomy, Ohio State University, 174 W. 18th Avenue, 
Columbus OH 43210, USA \\
$^3$Shaun.Cole@durham.ac.uk\\
$^4$S.J.Hatton@durham.ac.uk \\
$^5$dhw@astronomy.ohio-state.edu \\
$^6$C.S.Frenk@durham.ac.uk\\
}

\maketitle

\begin{abstract}
We present a comprehensive set of mock 2dF and SDSS galaxy redshift 
surveys constructed from a set of large, high-resolution cosmological 
\nbody simulations.  The radial selection functions and geometrical limits
of the catalogues mimic those of the genuine surveys. 
The catalogues span a wide range of cosmologies, 
including both open and flat universes.  In all the models the galaxy
distributions are biased so as to approximately reproduce the
observed galaxy correlation function on scales of $1$--$10 \hmpc$.
In some cases models with a variety of different biasing prescriptions
are included. All the mock catalogues are publically available at
\WWW. We expect these catalogues to be a valuable aid in the development
of the new algorithms and statistics that will be used to analyse the 2dF
and SDSS surveys when they are completed in the next few years. Mock
catalogues of the PSCZ survey of IRAS galaxies are also available at the
same WWW location.
\end{abstract}

\begin{keywords}
cosmology: theory -- large-scale structure of Universe -- galaxies: clustering 
\end{keywords}

\section{Introduction}
\label{sec:intro}

Our knowledge of large scale stucture in the Universe is going to change
dramatically as a result of the new generation of galaxy redshift surveys
now underway. The Anglo-Australian 2-degree Field (2dF) galaxy redshift
survey will measure redshifts for 250,000 galaxies selected from the APM
galaxy survey (Maddox \etal 1990), and the Sloan Digital Sky Survey (SDSS)
will obtain a redshift sample of one million galaxies.  These surveys will
be more than one order of magnitude larger than any existing survey and will
allow measurements of large-scale structure of unprecedented
accuracy and detail.  Precise estimates of the standard statistics that are
used to quantify large-scale structure (e.g., the galaxy correlation
function $\xi(r)$ and power spectrum $P(k)$) will be possible, and the
surveys will provide the first opportunity to examine more subtle
properties of the galaxy distribution.  Achieving these goals will require
the development of faster algorithms capable of dealing with the very large
numbers of galaxies involved, and the development of new statistical
measures.  To facilitate both of these tasks before the surveys are
complete will require synthetic data sets on which the techniques can be
developed and tested.

This paper presents an extensive set of mock 2dF and SDSS galaxy
catalogues.  These artificial galaxy redshift catalogues have been constructed
from a series of large, high-resolution cosmological \nbody
simulations.  The \nbody simulations span a wide range of cosmological
models, with varying values of the density parameter, $\Omega_0$, and the
cosmological constant, $\Lambda_0$, and with varying choices of the shape and
amplitude of the mass fluctuation power spectrum, $P(k)$.  For some
models several different catalogues have been produced, each employing
a different biasing algorithm to relate the galaxy distribution to the
underlying mass distribution. All the mock galaxy catalogues have
selection functions that mimic those expected for the real surveys.
The details of the construction of catalogues and their basic properties
are described here. The catalogues themselves can be obtained from \WWW .

The mock redshift catalogues are the principal scientific product
of this paper.  We expect to use them ourselves as we prepare for 
the analysis of large-scale structure in the 2dF and SDSS redshift surveys.
We are making them publically available in the hope that they will
be useful to other researchers, both inside and outside the two
collaborations.  Our illustrative plots also provide a qualitative
prediction of the structure expected in these redshift surveys
if the leading scenario for structure formation, based on Gaussian
primordial fluctuations and a universe dominated by cold dark matter,
is basically correct.  The mock catalogues have a number of limitations
(discussed in \S\ref{sec:limits} below) --- for example, the $345.6\hmpc$
simulation cubes are not as large as one might like, and we do not
model some of the detailed selection biases that will affect the real 
surveys, such as loss of members of close galaxy pairs because of a minimum
fibre separation.  The strength of this collection of catalogues is
that it covers a wide range of theoretically interesting cosmological
models in a systematic, homogeneous, and documented fashion.
We anticipate that these catalogues will be especially helpful to
researchers who want to test the discriminatory power of statistical
techniques that probe intermediate scale clustering ($\sim 1-100\hmpc$)
and/or to develop practical implementations of these techniques for
large data sets. Eventually, mock catalogues like these, or 
improved versions of them, will be a 
valuable tool for comparing the survey data against the predictions of 
cosmological theories. 

The cosmological models we have selected fall into two sets, which
we refer to as ``\COBE normalized'' and ``structure normalized'' 
(or ``cluster normalized'').
In the \COBE normalized models, the amplitude of the density fluctuations 
is set by the amplitude of the cosmic microwave
background temperature fluctuations measured by the \COBE satellite and 
extrapolated to smaller scales using standard assumptions. 
The shape of the spectrum of density fluctuations is fixed by
applying additional constraints on the age of the universe and the
baryon fraction.  The structure normalized models are, on the other hand, 
intended to produce approximately the observed abundance of
rich galaxy clusters, and all of them have the same shape for the density
fluctuation spectrum, chosen to be consistent with existing 
observations of large-scale structure.  Each set contains both open
($\Lambda_0=0$) and flat ($\Omega_0+\Lambda_0=1$) models with a range of
values of $\Omega_0$. Some of the models we consider come close to
satisfying simultaneously the \COBE and cluster abundance constraints. 
For each simulation we apply a ``biasing'' algorithm to select galaxies
from the \nbody particle distribution, choosing its parameters
so that the simulated galaxy population approximately reproduces 
the amplitude and slope of the observed
galaxy correlation function on scales $\sim 1-10\hmpc$.
For a few of the models we create multiple catalogues using several different
biasing schemes, so that the sensitivity of methods to the detailed
properties of biased galaxy formation can be investigated.
The \COBE normalized models arguably have a stronger theoretical motivation,
since they represent the predictions of models that assume
inflationary primordial fluctuations and cold dark matter with
the specified values of $\Omega_0$, $\Lambda_0$, $\Omegab$,
and the Hubble constant. Since
the structure normalized CDM models all have the same spectral shape
they are particularly useful for testing techniques
designed to measure $\Omega_0$ or $\Lambda_0$ independently of
an assumed shape of the primordial mass power spectrum.
We have presented analyses of aspects of 
these simulations elsewhere (\cite{ecf96}; Cole \etal 1997,
hereafter \cite{CWFR97}; \cite{hc98}).

The paper is structured in the following way.  The choice and
parameterization of the cosmological models is discussed in
Section~\ref{sec:models}.  Section~\ref{sec:sims} is a full description of
all the details of our \nbody simulations.  The construction of the
initial conditions and their evolution are described in
Sections~\ref{sec:init} and~\ref{sec:evolve}. The biasing prescriptions
are explained in Section~\ref{sec:bias}.  
The method by which the biased distributions
are converted into mock galaxy catalogues is presented in
Section~\ref{sec:cats}.  Our modelling of the survey geometries and
selection functions is detailed in Section~\ref{sec:select}.
Section~\ref{sec:illust} presents plots showing slices of the galaxy
distributions in a selection of the mock galaxy catalogues.  The
qualitative differences that are discernible in these distributions
and the processes that give rise to them are discussed.
In Section~\ref{sec:limits} we discuss the limitations of our 
approach.  
Section~\ref{sec:manual} gives instructions on how to obtain and
manipulate the mock galaxy catalogues.  We conclude in
Section~\ref{sec:conclude} .

\section{Cosmological Models}
\label{sec:models}

All our cosmological models are variants of the cold dark matter (CDM) 
scenario.
The functional form we adopt for the matter power spectrum is that given by
Bardeen et al. (\shortcite{BBKS}),
\begin{eqnarray}
   P(k) &\propto& \frac{k^n}  
   {\left[1 + 3.89q + (16.1q)^{2} + (5.46q)^3 + (6.71q)^4 \right]^{1/2}} 
\nonumber \\
&& \times \frac{\left[\ln(1+2.34q)\right]^2}{(2.34q)^2},
\end{eqnarray}
where $q=k/\Gamma$ and $k= 2 \pi /\lambda$ is the wavenumber in units
of $h {\rm Mpc}^{-1}$.  The index $n$ is the slope of the primordial
power spectrum, and in all but one case we adopt $n=1$, as predicted by
the simplest models of inflation.  Two further parameters complete the
description of the matter power spectrum.  These are the shape
parameter $\Gamma$ and the amplitude of the power spectrum, which we
specify through the value of $\sigma_8$, the linear theory rms fluctuation 
of the mass contained in 
spheres of radius $8 \hmpc$.  The background cosmological model in
which these fluctuations evolve is specified by the density parameter
$\Omega_0$ and the cosmological constant $\Lambda_0$, which we express in
units of $3H_0^2/c^2$, where $H_0$ is the present value of the Hubble
parameter.  Thus, with the exception of the one tilted model with $n
\ne 1$, our models are fully specified by the values of four
parameters, $\Omega_0$, $\Lambda_0$, $\sigma_8$ and $\Gamma$.  For each
of our twenty models, Table~\ref{tab:simpar} lists the values of these 
parameters along with other parameters that are described below.  The
names we have listed for the cosmological models are consistent with
the convention used in \cite{CWFR97}, but in addition we have included 
(in parentheses) some more descriptive names for the various 
$\Omega_0=1$ models.

\begin{table*}
\centering
\caption{Simulation Parameters: the first column gives 
the label of each of the cosmological models; alternative, more
descriptive names for the $\Omega_0=1$ models are given in
parentheses.  The following eight columns give the 
corresponding values of the density parameter
$\Omega_0$, cosmological constant $\Lambda_0$, Hubble parameter $h$,
age of the universe $t$, baryon content $\Omegab$, power
spectrum shape parameter $\Gamma$, and normalization $\sigma_8$
respectively. The final two columns give the
initial expansion factors, $a_{\rm i}$, and number of timesteps,
$N_{\rm steps}$, used in the \nbody simulation.}
\begin{center}
\begin{tabular}{llllllllllllllllr} \hline
\multicolumn{1}{l} {Model} &
\multicolumn{1}{l} {$\Omega_0$ } &
\multicolumn{1}{l} {$\Lambda_0$  } &
\multicolumn{1}{l} {h} &
\multicolumn{1}{l} {$t/{\rm Gyr}$ } &
\multicolumn{1}{l} {$\Omega_{\rm b}$ } &
\multicolumn{1}{l} {$\Gamma$  } &
\multicolumn{1}{l} {$\sigma_8$  } &
\multicolumn{1}{l} {$a_{\rm i}$  } &
\multicolumn{1}{l} {$N_{\rm steps}$} \\
\hline 
O3           & 0.3 & 0.0 & 0.65 & 12.2 & 0.030 & 0.172 & 0.5   & 0.15 &  93 \\ 
O4           & 0.4 & 0.0 & 0.65 & 11.7 & 0.030 & 0.234 & 0.75  & 0.1  & 168 \\ 
O5           & 0.5 & 0.0 & 0.6  & 12.3 & 0.035 & 0.27  & 0.9   & 0.08 & 254 \\ 
L1           & 0.1 & 0.9 & 0.9  & 13.9 & 0.015 & 0.076 & 0.7   & 0.15 & 150 \\
L2           & 0.2 & 0.8 & 0.75 & 14.0 & 0.022 & 0.131 & 0.9   & 0.12 & 220 \\ 
L3           & 0.3 & 0.7 & 0.65 & 14.5 & 0.030 & 0.172 & 1.05  & 0.101& 266 \\ 
L4           & 0.4 & 0.6 & 0.6  & 14.5 & 0.035 & 0.213 & 1.1   & 0.09 & 275 \\ 
L5           & 0.5 & 0.5 & 0.6  & 13.5 & 0.035 & 0.27  & 1.3   & 0.07 & 331 \\ 
O2S          & 0.2 & 0.0 & --   &  --  & --    & 0.25  & 1.44  & 0.028& 447 \\ 
O3S          & 0.3 & 0.0 & --   &  --  & --    & 0.25  & 1.13  & 0.050& 313 \\ 
O4S          & 0.4 & 0.0 & --   &  --  & --    & 0.25  & 0.95  & 0.073& 258 \\ 
O5S          & 0.5 & 0.0 & --   &  --  & --    & 0.25  & 0.83  & 0.096& 212 \\ 
L2S          & 0.2 & 0.8 & --   &  --  & --    & 0.25  & 1.44  & 0.057& 405 \\
L3S          & 0.3 & 0.7 & --   &  --  & --    & 0.25  & 1.13  & 0.080& 287 \\ 
L4S          & 0.4 & 0.6 & --   &  --  & --    & 0.25  & 0.95  & 0.102& 224 \\ 
L5S          & 0.5 & 0.5 & --   &  --  & --    & 0.25  & 0.83  & 0.122& 184 \\
E1\hphantom{S} (CCDM)    & 1.0 & 0.0 & 0.5  & 13.1 & --    & 0.5   & 1.35  & 0.061& 327 \\
E2\hphantom{S} (tilted)  & 1.0 & 0.0 & 0.5  & 13.1 & 0.05  & --    & 0.55  & 0.20 & 200 \\ 
E3S ($\tau$CDM) & 1.0 & 0.0 & --   & --   & --    & 0.25  & 0.55  & 0.21 & 103 \\ 
E4\hphantom{S} (SCDM)    & 1.0 & 0.0 & 0.5  & 13.1 & --    & 0.5   & 0.55  & 0.15 & 170 \\ 
\hline
\end{tabular} 
\end{center}
\end{table*}
\label{tab:simpar}

The \COBE-normalized set of models consists of an Einstein-de Sitter,
$\Omega=1$, model (labelled E1 or CCDM for \COBE normalized CDM), three
open models with $\Omega_0=0.3$, $0.4$ and~$0.5$ (labelled O3-O5) and
five flat models with $\Omega_0=0.1$-$0.5$ and $\Omega_0+\Lambda_0=1$
(labelled L1-L5).  We do not include \COBE normalized open models
with $\Omega_0=0.1$ or 0.2 because they are hopelessly inconsistent
with the observed abundance of rich galaxy clusters (\cite{CWFR97}).
For each of the open models we choose the value of
the Hubble parameter $h$\footnote{We use the convention that $h$ is
the value of the Hubble parameter in units of $100 \, \kms {\rm
Mpc^{-1}}$} that gives a universe of age $t \approx 12$Gyr,
\ie the largest value of $h$ that is consistent with standard globular
cluster age estimates (\cite{cdkk96,renzini96}; Salaris, Degl'Innocenti 
\& Weiss \shortcite{salaris96}).  For each of the low-$\Omega_0$
flat models we choose the value
of $h$ that gives $t \approx 14$Gyr.  For $\Omega_0=1$ we take
$h=0.5$.  Having chosen these values of $h$, we fix the baryon
fraction in these models using the constraint from primordial
nucleosynthesis of $\Omegab= 0.0125 h^{-2}$ (\cite{walker91}).  
We then use the
following expression for the shape parameter $\Gamma$,
\begin{equation}
\Gamma = \Omega_0 h \, \exp(-\Omegab -\Omegab/\Omega_0),
\label{eqn:shape}
\end{equation}
which approximately accounts for the effect of baryons on the transfer
function (\cite{sugiyama95})\footnote{The expression for $\Gamma$ which
we have adopted is from the original version of the Sugiyama (1995) 
paper and differs slightly
from the expression in the published version of that paper, which was modified
to improve its accuracy for high values of $\Omegab$.}.  
Finally, in each of these models the amplitude of the
density perturbations is set so as to be consistent with the \COBE
measurements of fluctuations in the cosmic microwave background 
(\cite{smoot92}).
Further details of these models can be found 
in \cite{CWFR97}, which examines the abundance 
of galaxy clusters in \COBE normalized CDM and presents some 
analysis of clustering of the mass distributions. 

For the set of structure normalized models, we adopt a fixed value of
$\Gamma=0.25$, as suggested by observations of the large-scale structure
traced by galaxies (\eg \cite{maddox90}).
The amplitude of the power spectrum we set
according to the formula, $\sigma_8 = 0.55 \Omega_0^{-0.6}$, which
results in an abundance of rich
galaxy clusters in good agreement with observations
(\cite{wef93}).  These models include the Einstein-de Sitter model
E3S (of which we have two realizations labelled E3S~A and E3S~B), 
a series of open models with $\Omega_0=0.2$-$0.5$ (labelled 
O2S-O5S), and a series of flat models with $\Omega_0=0.2$-$0.5$ and
$\Omega_0+\Lambda_0=1$ (labelled L2S-L5S).  Physically, these models
could be produced either by having $h=\Gamma/\Omega_0$ or by a change
from the standard model of the present energy density in relativistic
particles.  For example the E3S model is very similar to the $\tau$CDM model
of Jenkins \etal (\shortcite{virgo97}),
which is motivated by the decaying particle
model proposed by Bond \& Efstathiou (\shortcite{be91}).
The final model
listed in Table~\ref{tab:simpar}, E4 (SCDM), is the ``standard'' CDM model
($\Gamma=h=0.5$), normalized by the abundance of galaxy clusters.

We consider one further model that falls into both the
\COBE and structure normalized categories.  This is the tilted 
Einstein-de Sitter model, E2 (tilted).  For this model, the above 
constraints have been
applied in relating the baryon fraction $\Omegab$, the Hubble
parameter $h$, and the shape parameter $\Gamma$, 
but in addition the slope 
$n$ of the primordial power spectrum has been adjusted to match 
\COBE observations at large scales while simultaneously achieving 
$\sigma_8=0.55$, as required for consistency with the observed cluster 
abundances.  
This procedure results in a tilted primordial spectrum with
$n=0.803$ and a transfer function with $\Gamma=0.4506$ as given by
equation~\ref{eqn:shape}.  
In normalizing to the \COBE observations, we have included a
gravitational wave contribution as predicted by the power-law model of
inflation. For our model gravitational waves contribute  approximately
$55$\% of the rms temperature fluctuations on the scales probed by COBE.

\section{N-body Simulations}
\label{sec:sims}

We now describe how the initial conditions of our simulations were
set up, how the simulated mass distribution was
propagated to the present day, and how the particles labelled
as galaxies were selected.

\subsection{Initial conditions}
\label{sec:init}

Before imposing the desired density perturbations, we set up a
`uniform' distribution using the technique described by 
White (\shortcite{white94}) and
Baugh, Gazta\~naga \& Efstathiou (\shortcite{baugh95})
to generate a particle distribution with a
`glass' configuration.  This was achieved by first randomly placing
$192^3$ particles throughout the simulation box and then evolving
this distribution with the \nbody code, but with the sign of the
gravitational forces reversed.  We used large timesteps which were
approximately logarithmically spaced in expansion factor and evolved
the distribution until the gravitational forces on all particles
practically vanished.  With this approach, the initial particle
distribution is not regular, but the small random fluctuations in the
particle density field do not seed the growth of spurious structures.
Simulations with glass and grid initial conditions have been found to
give very similar statistical results once they are evolved into the
nonlinear regime (\cite{white94,baugh95}), but the simulations with
glass initial conditions have the advantage that they do not retain an
unseemly grid signature in uncollapsed regions.

Each of the simulations was of a periodic box of side $345.6 \hmpc$
($192 \times 1.8 \hmpc$).   For each, we created a Gaussian random
density field on a $192^3$ grid, using the same Fourier phases from
one model to the next, but varying the mode amplitudes according to
the model power spectrum.  We applied the Zel'dovich approximation to
this density field to compute displacements and peculiar velocities at
each grid point.  We then displaced each particle from its `glass'
position according to the displacements interpolated from the
grid values.  The initial expansion factors of the simulations $a_{\rm
i}$, listed in Table~\ref{tab:simpar}, were determined by setting the
amplitude of the initial power spectrum at the Nyquist frequency of
the particle grid to be $0.3^2$ times that for an equivalent Poisson
distribution of particles.  Thus $P_{\rm initial}(k_{\rm N}) =
0.3^2/\bar n$, where $\bar n$ is the mean particle density and the
Nyquist frequency is $k_{\rm N}=\pi \bar n^{1/3} = (2 \pi/3.6)h\;{\rm
Mpc}^{-1}$.  The residual power in the glass configuration is only
0.5\% of that in a Poisson distribution at the Nyquist frequency and
drops very rapidly at longer wavelengths (see figure~A2 of
\cite{baugh95}).  This choice is safely in the regime where (a) the
initial density fluctuations are large compared to those present in
the glass, but (b) the Zel'dovich
approximation remains accurate.  In particular, no shell-crossing has
occurred.

\subsection{Evolution}
\label{sec:evolve}

We evolved the simulations using a modified version of Hugh 
Couchman's Adaptive Particle-Particle-Particle-Mesh (AP$^3$M, 
\cite{ap3m91}) \nbody code. 
We set the softening parameter of AP$^3$M's triangular-shaped cloud
force law to $\eta=270\hkpc$, $15\%$ of the grid spacing.  
The softening scale is fixed in comoving co-ordinates.
This choice corresponds approximately to a gravitational softening
length $\epsilon=\eta/3=90\hkpc$ for a Plummer force law,
and we adopt $\epsilon$ as our nominal force resolution.
The size of the timestep $\Delta a$ was
chosen so that the following two constraints were satisfied throughout the
evolution  of the particle distribution.  First, 
the rms displacement of particles in one timestep was less
than $\eta/4$.  Second, the fastest moving particle moved less than
$\eta$ in one timestep.  Initially these two constraints are comparable,
but at late times the latter constraint is more stringent, particularly
in low $\Omega_0$ simulations.  We monitored energy conservation
using the Layzer-Irvine equation (equation 12b of \cite{edwf85}) 
and found that for this choice of timestep, 
energy conservation with a fractional accuracy of better than $0.3$\%
was achieved.
We also tested the inaccuracy incurred by these choices of starting 
amplitude and timestep by comparing the final particle positions with
two additional versions of the E1, $\Omega_0=1$ simulations, which were 
run starting from a 
fluctuation amplitude a factor of two lower and using timesteps 
a factor of two smaller.
In each case we found that the final particle positions agreed very 
accurately, with rms differences of less than $\epsilon$.
More importantly, the correlation functions
of the particle distributions in all cases were indistinguishable at scales
larger than $\epsilon=90 \hkpc$.  Thus, the statistical clustering
properties of these simulations have a resolution that is limited
by the particle mass and force softening 
and not by the choice of timestep or starting redshift.

\subsection{Biasing}
\label{sec:bias}

In this section we describe the methods we use to select the particles
we label as galaxies from the distributions of mass produced in the
\nbody simulations.  It is unlikely that galaxies are unbiased tracers
of the underlying mass distribution.  This would only occur if the
ability to form a galaxy were independent of the properties of the
surrounding density field, so that each mass particle no matter
where it resided was equally likely to be associated with a galaxy. 
Simple, physically motivated models such as the high peaks model 
(\cite{DEFW,BBKS}) illustrate how a dependence of galaxy formation on the 
properties of the local density field can make the galaxy distribution
more strongly clustered than the underlying mass distribution.  This effect
can be quantified in terms of a bias factor $b_r= \sigmag_r/\sigmam_r$,
relating the fractional rms fluctuation in the number of galaxies
in spheres of radius $r \hmpc$ to the corresponding 
variation in the mass.

Observational evidence for bias is presented by Peacock \& Dodds 
(\shortcite{PD94}).  They assume a simple, constant linear bias model in which 
a perturbation in the mass distribution is accompanied by an amplified
perturbation in the galaxy distribution, $\delta_{\rm gal} = b
\delta_{\rm mass}$.  They find that the power spectra of differently
selected galaxy samples require a bias relative to the power spectrum of
IRAS galaxies, $b/b_{\rm IRAS}$, of
4.5:1.9:1.3, for Abell clusters, radio galaxies and optically selected
galaxies respectively.  Since a relative bias 
exists between any two of these differently selected samples, it 
seems natural to 
assume that all galaxy samples will be subject to some degree of bias.
We note that bias is also important in interpreting the estimates of
the mass-to-light ratio of galaxies in clusters.  These have been used
in conjunction with estimates of the galaxy luminosity function to infer
$\Omega_0 \approx 0.2$ (\eg Carlberg, Yee \& Ellingson 1997).  
This inference assumes that galaxies are
unbiased tracers of the mass distribution.  If galaxies form preferentially
in proto-cluster environments, then this estimate translates to
$\Omega_0/B \approx 0.2$, where $B$ is the factor by which the 
efficiency of galaxy formation is enhanced in regions destined to 
become clusters, relative to the field. 

Since the physics of galaxy formation is very complex, it is not yet
possible to determine the function that relates the probability of
forming a galaxy to the properties of the mass density field,  
though first steps towards this goal have been taken using
cosmological simulations with gas dynamics 
(\cite{cen92,khw92,summers95,fews96,j96}).
For this reason we take the approach of defining our biasing algorithm
in terms of a simple parametric function.  Then for each cosmological
model we constrain the values of the function's parameters using
estimates of observed small and intermediate scale galaxy clustering.  
For a subset
of the cosmological models we repeat this procedure for a
variety of different biasing algorithms.  This enables the extent to
which the properties of the catalogues depend on the arbitrary choice
of the adopted biasing algorithm to be quantified.

\begin{table*}
\centering
\caption{Bias Model Parameters: For the three selected cosmological models
we list the parameter values required in each of the six bias models.
The resulting galaxy correlation functions 
are compared in Fig.~\ref{fig:xi3}. }
\begin{center}
\begin{tabular}{lrrrrrrrrrrrrrrr} \hline
\multicolumn{1}{l} {Identifier} &
\multicolumn{2}{c} {Model 1} &
\multicolumn{2}{c} {Model 2 } &
\multicolumn{1}{c} {Model 3 } &
\multicolumn{1}{c} {Model 4 } &
\multicolumn{1}{c} {Model 5 } &
\multicolumn{2}{c} {Model 6 } \\
 & $\alpha_{\rm i}$ & $\beta_{\rm i}$ & $\alpha_{\rm f}$  & $\beta_{\rm f}$  & $\nu_{\rm p}$  & $\rho_{\rm T}$  & $\alpha$   & $\alpha_{\rm f}$  & $\beta_{\rm f}$  \\
\hline 
O4S            & 3.60 &  -9.05 & 2.17 & -1.31  & --    & $<$19.7  & -0.02 &3.96 &-2.69\\
L3S            & 2.55 & -17.75 & 0.15 & -0.06  & --    & $<$15.5  & -0.13 &7.11 &-7.15\\
E3S            & 1.10 &  -0.56 & 1.26 & -0.51  & 1.005 & $>$0.98  & 0.56 &2.98 &-1.25\\
\hline
\end{tabular} 
\end{center}
\end{table*}

For optically selected galaxies in the APM survey, $\sigma_8^{gal}=0.96$
(\cite{apmIII}). Many of our simulations have $\sigmam_8> 0.96$ and
therefore require an anti-bias ($b<1$) on the $8\hmpc$ scale.  Anti-bias
seems less physically motivated than bias because it requires negative
feedback processes to suppress galaxy formation in high-density regions.
Such an anti-correlation, however, might be produced even if the production
rate of galaxies in proto-clusters is higher than in low-density regions,
so long as galaxy merging in the proto-clusters is sufficiently
efficient to suppress the overall number of galaxies in clusters.

All the biasing schemes we consider are local, in the sense that the 
probability of a 
mass particle being selected as a galaxy is a function only of the
neighbouring density field, \eg the density field smoothed on a scale
$3\hmpc$.  Such models have the property that on scales (in the linear
regime) that are much larger than that defining the neighbourhood they
produce a constant, scale independent bias (\cite{sw98}).  A derivation of
an expression for this asymptotic bias is given in Section~\ref{sec:asymp}
below.  Our algorithms include both Lagrangian models, in which the
selection probability is a function of the initial density field, 
and Eulerian models, in which the probability is a function of the final mass 
density field.  For a consideration of the
differences between these approaches, see Mann, Peacock \& Heavens
(\shortcite{MPE97}).

We use six different prescriptions for creating the biased galaxy
samples.  All of them involve defining a probability field from
either the initial or the final density distribution, and then Poisson
sampling the simulation particles using this field to define the
selection probability.   The probability is normalized such that a mean
of $128^3$ out of our original $192^3$ particles are selected.  This
corresponds to a galaxy number density $\bar n_{\rm g} \approx 0.05 h^3{\rm
Mpc^{-3}}$, which approximately equals that of galaxies brighter than
$L_\star/80 $.  Although this density is less than that of the original
simulation, occasionally the bias may demand that in certain regions 
there is a greater galaxy density than the original particle density.
The Poisson sampling achieves this by allowing some particles to 
be selected more than once. This double sampling is generally rare but
can occur in the highly biased models.
The functions defining the selection probability
have one or two free parameters.  In the case of those with just one
free parameter, we fix its value by demanding that $\sigmag_8 = 0.96$, in
agreement with the value estimated from the APM galaxy survey.  The
models with two parameters ($\alpha$ and $\beta$)
enable us to control both the amplitude of
galaxy clustering on large scales and, to some extent, the slope of the
galaxy correlation function on small scales.  We set their parameters
by attempting to match simultaneously the observed variance of the
galaxy density field in cubic cells of $5$ and~$20\hmpc$ on a
side.  These, we take to be $\sigma_{\rm cell \, 5}=2.0$ and 
$\sigma_{\rm cell \, 20}=0.67$, the values we have obtained from 
the power spectrum shape estimated for APM galaxies by Baugh \& Efstathiou 
(\shortcite{BE94}), scaling its amplitude for consistency with the more recent
estimate of $\sigmag_8 = 0.96$.  In some cases, where, for instance,
the small scale mass correlation function is very much steeper than
the observed galaxy correlation, it does not prove possible to
simultaneously satisfy these two constraints.  For computational simplicity
and to avoid any ambiguity, we choose, in all cases, to fit the observed
values by minimizing the cost function
\begin{eqnarray}
  C(\alpha,\beta) &=& \left( \frac{(\sigma_{\rm cell \, 20}-0.67)}{0.67} 
      \right)^2
    + \left( \frac{(\sigma_{\rm cell\, 5}-2.0)}{2.0} \right)^2 \cr \nonumber
    &+& \epsilon_{\rm c} (\alpha^2 + \beta^2) .
\end{eqnarray}
The third term has $\epsilon_{\rm c} = 4 \times 10^{-7}$ and is included to 
avoid extremely large values of $|\alpha|$ and $|\beta|$ being
selected for very little improvement in the values of
$\sigma_{\rm cell \, 5}$ and $\sigma_{\rm cell \, 20}$.

\subsubsection{Biasing algorithms}

\begin{figure*}
\centering
\centerline{\epsfxsize= 15.5 truecm 
\epsfbox[0 50 574 740]{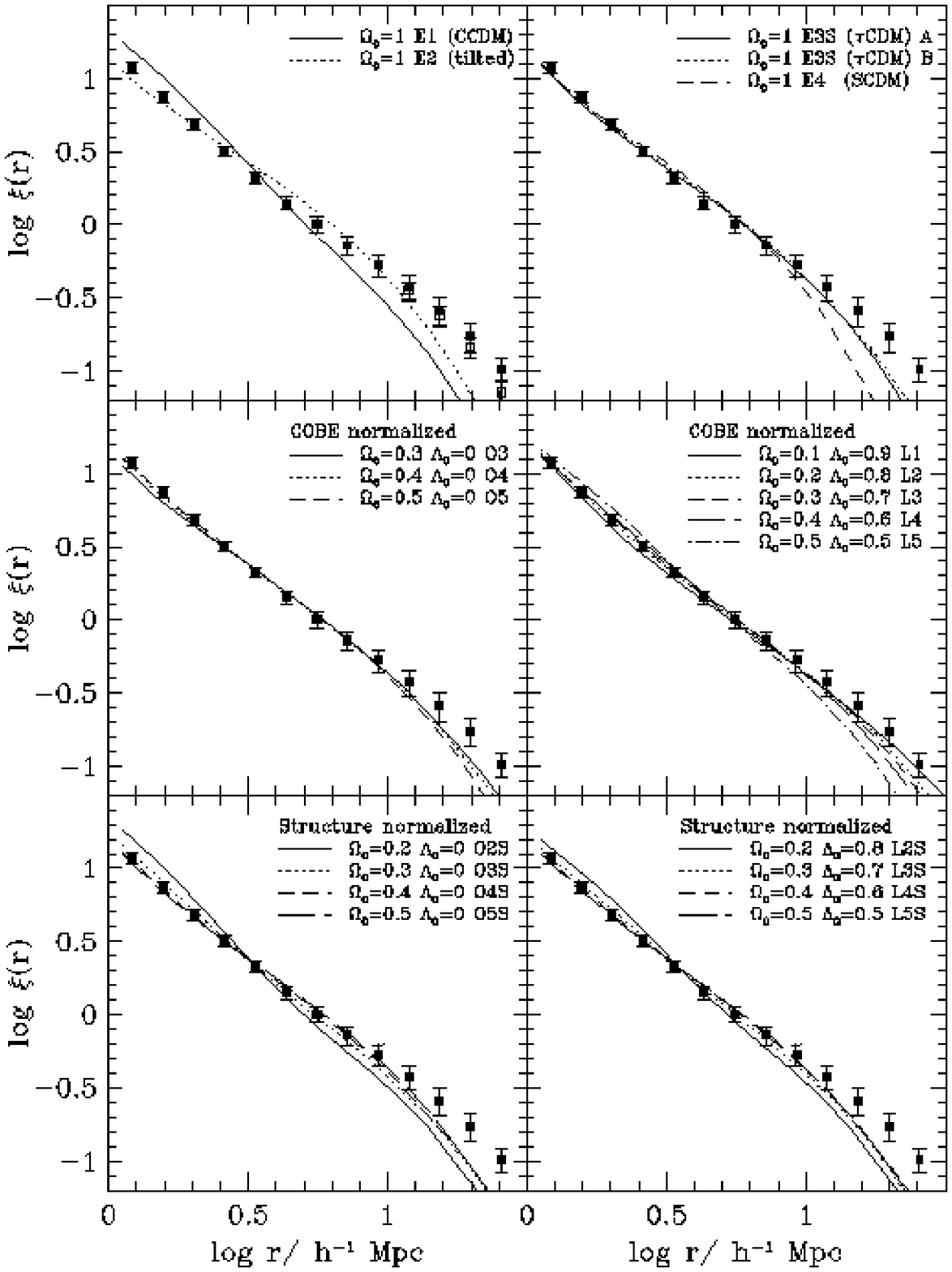}}
\caption{The galaxy correlation functions, $\xi(r)$,  for each of our 
cosmological models when biased using bias model~1.
Each of the lines corresponds to a different cosmological model as indicated
on the legend.  The solid data points are the same on each panel and are an
estimate of the galaxy correlation function from the APM survey 
(Baugh 1996). The open data points, shown on the first panel, show an 
alternative estimate of the APM correlation function obtained by Fourier
transforming the Baugh \& Efstathiou (1993) estimate of the APM power 
spectrum.
}
\label{fig:xi6_1}
\end{figure*}

\begin{figure*}
\centering
\centerline{\epsfxsize= 15.5 truecm 
\epsfbox[0 50 574 740]{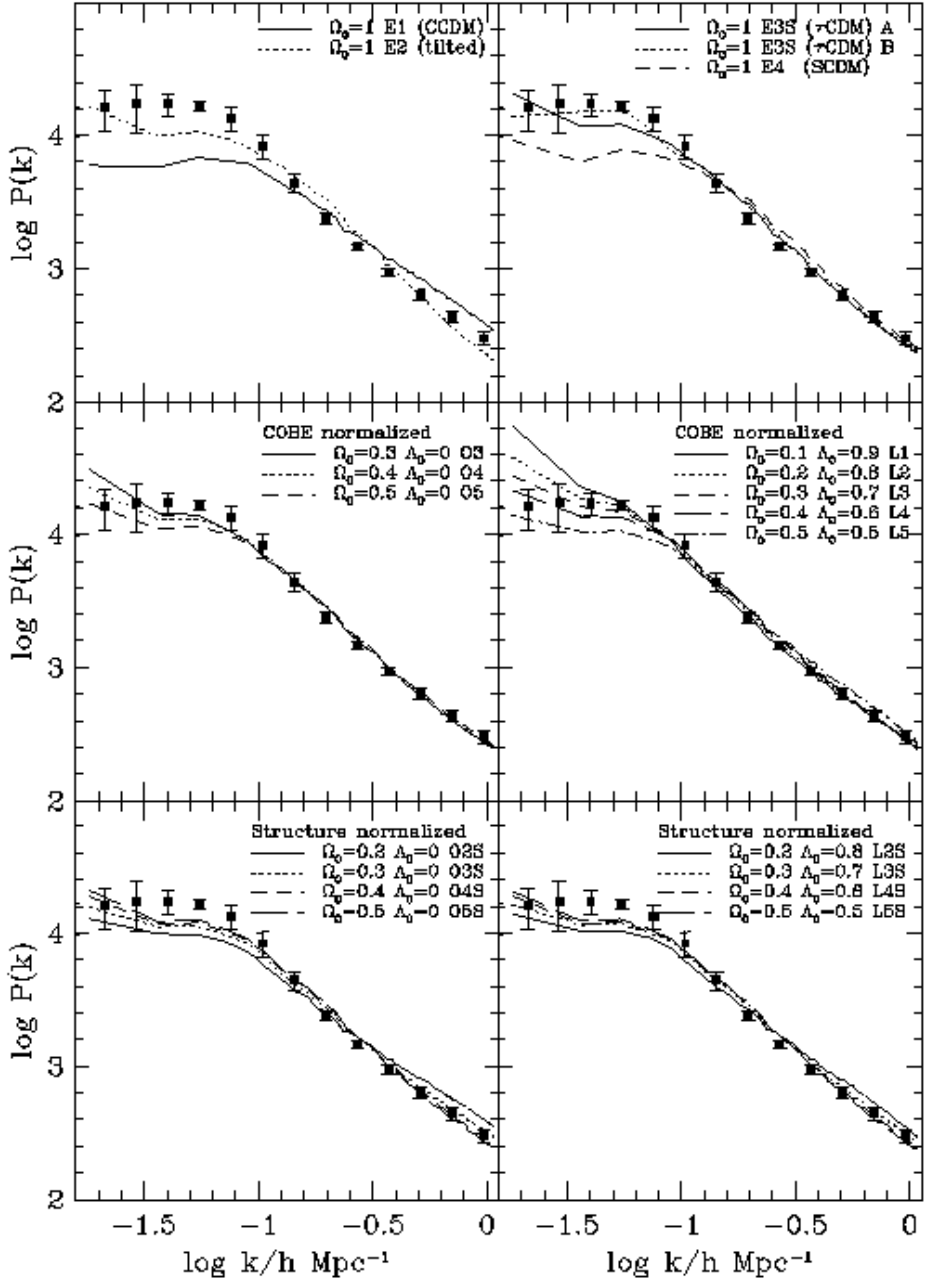}}
\caption{The galaxy power spectrum, $P(k)$, for each of our 
cosmological models when biased using bias model~1.
Each of the lines corresponds to a different cosmological model as indicated
on the legend.  The data points are the same on each panel and are an
estimate of the galaxy power spectrum from the APM survey 
(Baugh \& Efstathiou 1993).
}
\label{fig:pk6_1}
\end{figure*}

\begin{figure}
\centering
\centerline{\epsfysize= 20.9 truecm 
\epsfbox[35 66 300 699]{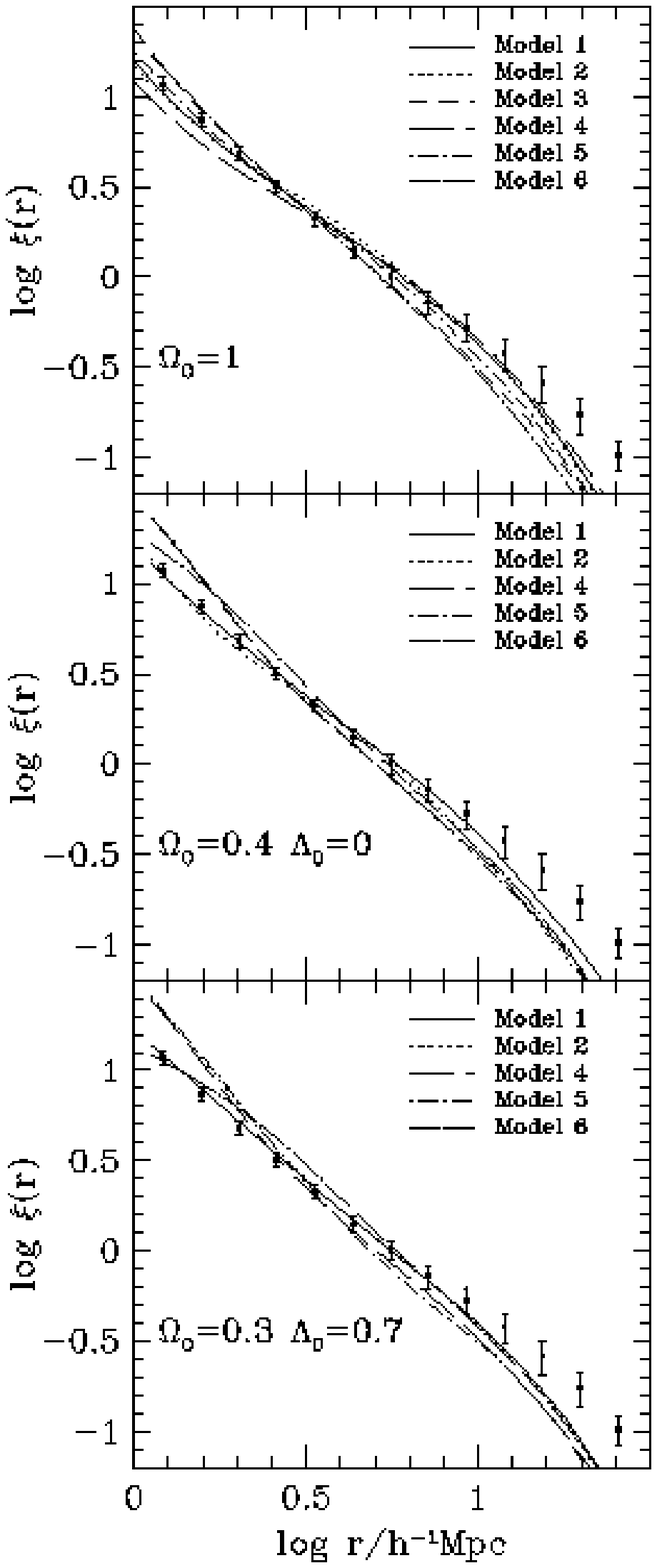}}
\caption{For three selected, structure normalized cosmological
models (E3S,O4S and~L3S), we show the galaxy correlation functions that
result from each of the bias models.  Note that both of the $\Omega_0<1$ models
require anti-bias and therefore cannot be biased using the peaks 
bias model~3.  The line types corresponding to each of the bias models are
indicated on the legend.  The data points again show the estimate of the
galaxy correlation function from the APM survey.}
\label{fig:xi3}
\end{figure}

Here we define the selection probability functions, $P(\nu)$,
which define each of our biasing algorithms.  The resulting biased 
galaxy correlation functions, $\xi(r)$, and power spectra, $P(k)$, are
shown in Figs.~\ref{fig:xi6_1}, \ref{fig:pk6_1}, and~\ref{fig:xi3} 
and discussed below. The biasing algorithm that
we apply to all of the cosmological models is model~1; the other biasing models
are used only to create additional mock catalogues for the O4S, L3S,
and E3S simulations.

\begin{enumerate}

\item[{\bf Model 1: }] This model bases the selection probability 
on the value of
the smoothed initial density.  The initial density field is smoothed
with a Gaussian of width $R_{\rm S} = 3 \hmpc$ (in $\exp(-r^2/2R_S^2)$) 
to define a smoothed density 
field $\rho_{\rm S} (\r) $ at the initial particle position.  A
dimensionless variable $\nu$ is defined as 
$\nu(\r)~=~\delta_{\rm S}(\r)/\sigma_{\rm S}$, where the density perturbation
$\delta_{\rm S}(\r)~=~(\rho_{\rm S}(\r)~-~\bar \rho)~/~\bar \rho $ ,
and $\sigma^2_{\rm S}= \langle \vert \delta_{\rm S} \vert^2 \rangle $.
We then adopt 
\begin{equation}
P(\nu) \propto \cases{ \exp(\alpha\nu+\beta\nu^{3/2}) &if $\nu \ge 0$ \cr
                       \exp(\alpha\nu)                &if  $\nu \le 0$ \cr },
\label{bias1}
\end{equation}

\noindent as the selection probability.  The model has two free parameters 
$\alpha$ and $\beta$.  This choice of functional form is essentially
selected for its simplicity.  Its exponential form ensures that the
probability cannot be negative.  The dependence on $\beta$ for $\nu>0$
enables the selection probability to be enhanced ($\beta>0$) or suppressed
($\beta<0$) in the densest regions.  It is this property which gives
some control over the slope of the small scale correlation function.
The choice of $\nu^{3/2}$ is such that the probability converges
when integrated over a Gaussian distribution of $\nu$.

\item[{\bf Model 2: }] For this model the same functional form (equation \ref{bias1})
is used to define the selection probability, but this time the variable 
$\nu$ is defined in terms of the smoothed {\it final} density field around
each particle.  Again, a Gaussian smoothing with $R_{\rm S} = 3 \hmpc$
is adopted.

\item[{\bf Model 3: }] This is the standard high peaks model of 
Bardeen \etal (\shortcite{BBKS}).
Their results are used to predict the number of 
peaks of amplitude $\nu>\nu_{\rm p}$ defined on the scale of a galaxy 
as a function of the density smoothed on a larger scale that {\it is} 
resolvable in our simulation.  In this case we choose the larger scale
to be defined by applying a sharp cutoff to the power at a wavelength
$\lambda \lsim 4\Mpc$, which is quite well resolved in the
initial conditions of the simulation. We define the galaxy mass scale
by a Gaussian smoothing with $R_{\rm S}=0.54\Mpc$ 
as adopted by White \etal (\shortcite{wfde87}).  Here the model
parameter is $\nu_{\rm p}$.  An unavoidable property of 
assuming that galaxies form in peaks of the density field
is that they are more clustered than the mass distribution ($b>1$).
Thus this method cannot be applied in cases where an anti-bias is required.

\item[{\bf Model 4: }]
In this model a sharp cut-off is applied to the final smoothed density field, 
so that galaxies are entirely prohibited from forming in very 
underdense regions, but have an equal chance of forming wherever 
the overdensity rises above a certain threshold, $\rho_{\rm T}$. Thus 
\begin{equation}
	P(\nu) \propto \cases{ 1 &if $\rho (\r) \ge \rho_{\rm T}$ \cr
                               0 &if  $\rho (\r) \le \rho_{\rm T}$ \cr } .
\end{equation}
This is the case if a bias greater than unity is required.  
For an anti-bias, the conditions are reversed and galaxies 
are prohibited from forming in the very densest regions.
Note that this prescription for producing anti-bias seems quite
unphysical, as it implies that the highest mass density regions
have no galaxies at all.

\item[{\bf Model 5: }] As in model~2 the selection probability is defined
in terms of the smoothed final density, but this time the functional
form adopted is a power law,
\begin{equation}
P(\nu) \propto \nu^{\alpha} .
\end{equation}
Here a positive value of the parameter, $\alpha$, will induce a
bias ($b>1$) and a negative value an anti-bias ($b<1$).
The bias inferred by Cen \& Ostriker (\shortcite{cen93}) from 
their hydrodynamic cosmological simulations has roughly this
form, with $\alpha \approx 1.5$.

\item[{\bf Model 6: }] This algorithm is a variation of model~2 and again
uses the formula (\ref{bias1}), but with a different definition of
the overdensity parameter $\nu$.  Instead of smoothing on a fixed scale of 
$3\hmpc$, the distribution was adaptively smoothed by setting the density
at the position of each particle, $\rho \propto 1/r^3_{10}$,
where $r_{10}$ is the distance to the 10th nearest neighbour of that
particle.

\end{enumerate}

The various galaxy correlation functions and power spectra resulting from
applying biasing model~1 to each of our cosmological simulations are shown
in Fig.~\ref{fig:xi6_1} and Fig.~\ref{fig:pk6_1} respectively.  The solid
data points show the estimates of the galaxy correlation function,
$\xi(r)$, (\cite{baugh96}) and power spectrum, $P(k)$, (\cite{baugh93}) of
APM galaxies, scaled in amplitude to match the updated estimate of
$\sigmag_8=0.96$ for the APM survey (\cite{apmIII}). The data points
plotted as open symbols on the top left panel of Fig.~\ref{fig:xi6_1} show
the APM correlation function as estimated from the Fourier transform of the
estimated APM power spectrum. There is a slight difference between this and
the direct estimate at large separations, which arises because both
$\xi(r)$ and $P(k)$ are estimated using non-linear inversions of the
measured angular correlation function.  The difference is an indication of
one of the systematic errors involved in 
estimating $\xi(r)$ on large scales.

In general the two parameter biasing
model is successful in matching both the amplitude and the shape of
the galaxy correlation function on scales of 
$1$--$10\hmpc$, as can clearly be seen in Fig.~\ref{fig:xi6_1}.  For a few 
cases, such as E1, O2S, and~L2S, which have high
values of $\sigmam_8$ and consequently steep non-linear mass
correlation functions, the bias model cannot  reduce the
slope of the correlation function enough 
to accurately match the observed value.
The behaviour of the correlation functions on large scales reflects
each model's value of the power spectrum shape parameter $\Gamma$.
The APM data, if fitted with a $\Gamma$-model, prefer
$\Gamma=0.15$--$0.2$ (\eg \cite{ESM90}), so even our structure normalized,
$\Gamma=0.25$ models fall short of the amount of large-scale power
evident in the APM correlation function.   
This short-fall is also exaggerated by a
statistical fluctuation in our simulation initial conditions.
As can be seen in the top-righthand panel of
Fig.~\ref{fig:pk6_1}, the first realization (A) of model E3S has less
power on scales $0.03\lsim k \lsim0.06 \, h $Mpc$^{-1}$ than the second
realization (B) of the same model.  This downward fluctuation in the
power is present in all the other cosmological models, since all the 
initial density fields were generated from the same basic Gaussian
random field but with expected mean amplitudes rescaled according to
the model power spectrum.  We also note that the longest wavelength
modes, with $k =0.018 \, h $Mpc$^{-1}$, are noisy because of the 
small number of such modes contained in the simulation box.
The comparison of model and APM galaxy power spectra on
small scales (high $k$) is in accord with the small scale behaviour of the
correlation functions.

The manner in which the galaxy clustering statistics vary with the
form of the  biasing is illustrated in 
Fig.~\ref{fig:xi3}. 
The 1-parameter bias models (models~3, 4 and~5) do not have the flexibility
to control both the amplitude and slope of the galaxy correlation function.
Thus, in general, these models do not match the APM galaxy 
correlation function over a wide range of scales.  In particular, the
galaxy correlation functions of the three 
models selected for Fig.~\ref{fig:xi3} 
are steeper than the correlation function of APM galaxies, 
reflecting the steepness of the underlying mass correlation functions.
The $3\hmpc$ filter used in bias model~2 smooths over the
structure of groups and clusters in the {\it final} density field. As a result,
the small-scale slope of the galaxy correlation function ends up being
insensitive to the bias model parameters in this case. 
In model~6, on the other hand, the use of an adaptive smoothing results 
in better resolution on the scale of groups and clusters. In some cases
this is enough to enable the required
adjustments to the slope of the correlation function on small scales.

\subsubsection{The asymptotic bias}
\label{sec:asymp}
In general all the biasing algorithms discussed above give rise to 
a bias that is scale dependent.  However, since these biasing algorithms 
only depend on 
local properties of the density field, the bias should tend to a
constant on large scales.  Where the selection probability is a function
of the initial density field, the value of this asymptotic bias can be
computed analytically.  The probability that a mass particle is
selected as a galaxy is taken to be $P(\nu)$, where $\nu$ is the
amplitude of the initial density fluctuation in units of the rms,
$\sigma_{\rm s}$.   The normalization of $P(\nu)$ is determined by the
integral over the Gaussian distribution of initial density
fluctuations,
\begin{equation}
	\frac{1}{\sqrt{2\pi}} \int P(\nu) \, e^{-\nu^2/2}  \, d\nu = 1 .
\end{equation}
The density of galaxies selected in a region in which a large
scale perturbation $\Delta$ is added will be given by
\begin{equation}
 \rho_{\rm gal} = \bar \rho_{\rm gal} \, 
		\frac{(1+\Delta) }{\sqrt{2\pi}} \int P(\nu^\prime) \, 
                e^{-\nu^2/2} \,d\nu ,
\end{equation}
where $\nu^\prime = \nu + \Delta/\sigma_{\rm s} $.
A first order series expansion of $P(\nu)$ yields
\begin{equation}
P(\nu^\prime)= P(\nu) 
\left (1 + \frac{d\ln P}{d\nu} \, \frac{\Delta}{\sigma_{\rm s}} \right ) .
\end{equation}
Hence
\begin{equation}
 \frac{\rho_{\rm gal}}{ \bar \rho_{\rm gal} } =
	\frac{(1+\Delta) }{\sqrt{2\pi}} 	\int P(\nu) 
\left (1 + \frac{d\ln P}{d\nu} \frac{\Delta}{\sigma_{\rm s}} \right) 
 e^{-\nu^2/2}  d\nu,
\end{equation}
which simplifies to
\begin{equation}
 \frac{\rho_{\rm gal}}{ \bar \rho_{\rm gal} } =
(1+\Delta) 
	\left( 1+ \frac{\Delta}{\sqrt{2\pi}\sigma_{\rm s}} \
\int \frac{dP}{d\nu} \, e^{-\nu^2/2}  \, d\nu  \right).
\end{equation}
We can thus define an asymptotic bias factor, \ie the ratio of the galaxy to 
the mass perturbations on large scales, as
\begin{equation}
  b_{\rm asymp} =  1 + \frac{1}{\sqrt{2\pi} \sigma_{\rm s}}
\int \frac{dP}{d\nu} \, e^{-\nu^2/2}  \, d\nu .
\end{equation}

\begin{figure*}
\centering
\centerline{\epsfxsize= 14 truecm 
\epsfbox{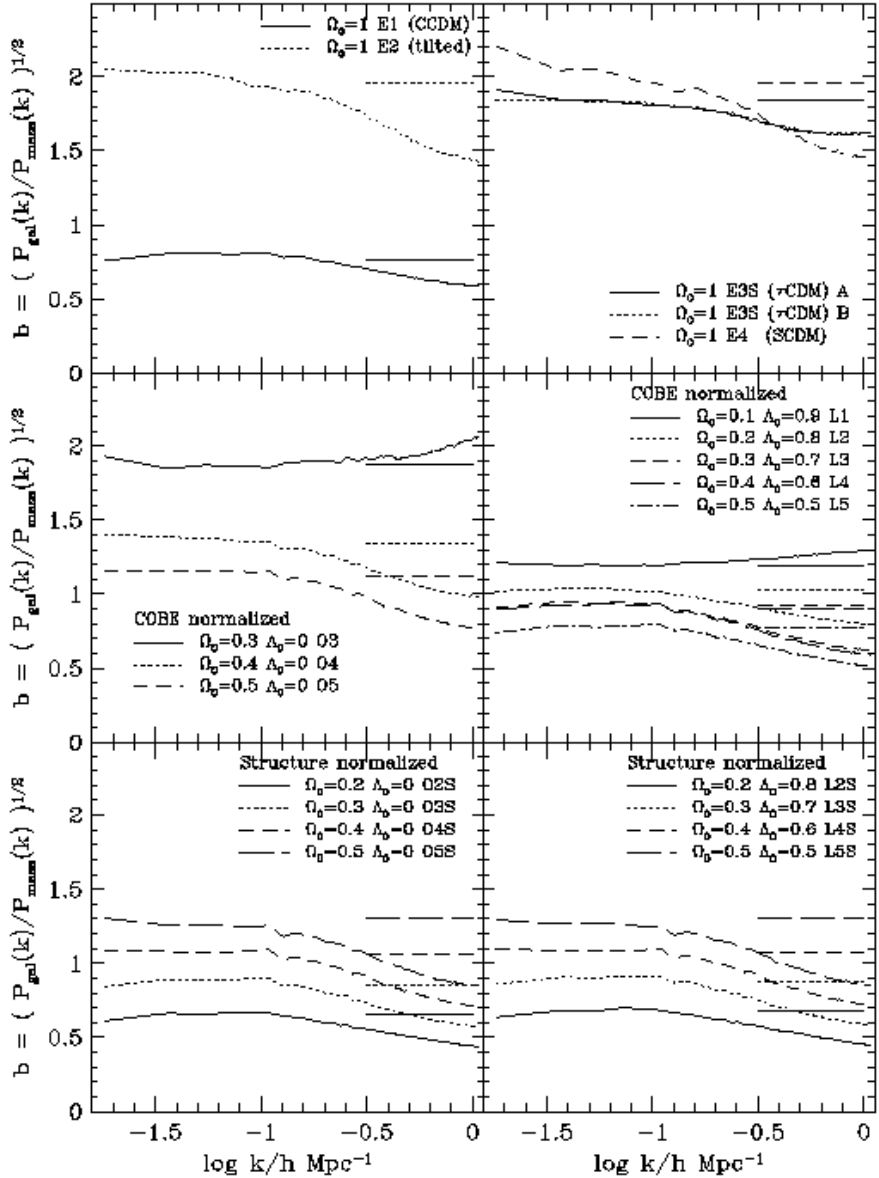}}
\caption{The scale-dependent bias, $b(k) = P_{\rm gal}(k) / P_{\rm mass}(k)$,
 for each of our cosmological models when biased using bias model~1.
Each of the lines corresponds to a different cosmological model as indicated
on the Figure.  To the right of each panel we show the value of the expected 
asymptotic bias on large scales, as explained in Section \ref{sec:asymp}.}
\label{fig:bias6_1}
\end{figure*}

This result is compared to the bias estimated from the simulations
in Fig.~\ref{fig:bias6_1}.  The figure clearly shows that the 
bias does indeed tend towards its asymptotic value, as calculated 
above, on large scales.  

\section{Mock Catalogues}
\label{sec:cats}

In the previous section we described the procedure by which we create
a galaxy distribution within each simulation cube. We now describe how these
are manipulated and sampled to create the mock galaxy catalogues.
It should be noted that we do not attempt to mimic the imperfections that
will inevitably be present in the genuine catalogues, \eg,
Galactic extinction, excluded
regions around bright stars, or missing members of galaxy pairs
separated by less than the minimum fibre spacing.  
Our goal is instead to create idealized
catalogues with the expected redshift distributions and geometrical properties
of the genuine surveys.
We anticipate that members of the 2dF and SDSS collaborations
will create a few mock catalogues that incorporate the finer details
of the survey properties.

\subsection{Survey geometry}
\label{sec:geometry}

\begin{figure*}
\centering
\centerline{\epsfxsize= 14 truecm 
\epsfbox[0 270 520 530]{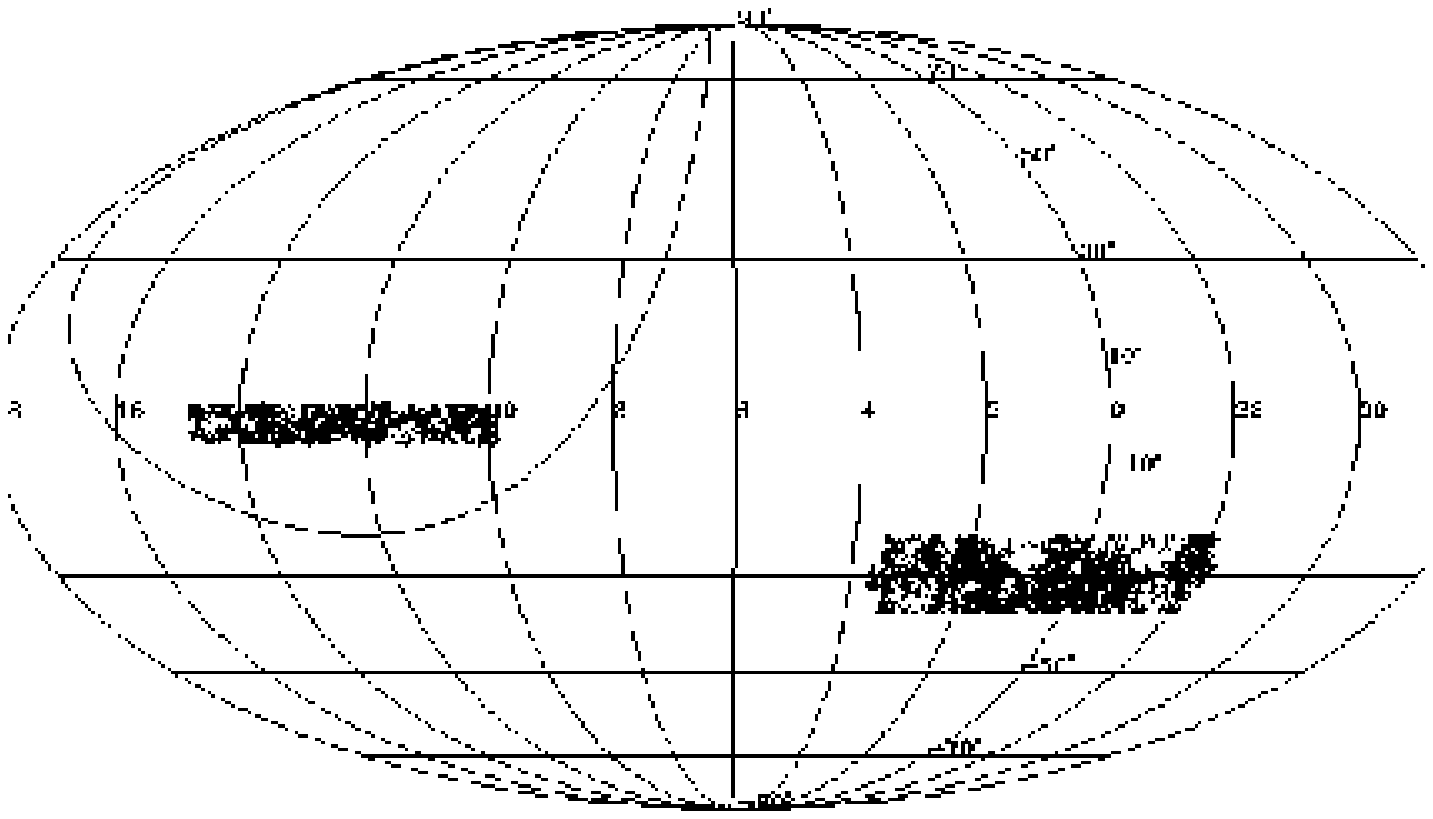}}
\caption{An equal area (Mollweide) projection of the whole sky 
showing the regions
covered by the 2dF and SDSS galaxy redshift surveys.  The regions covered by
the 2dF survey are indicated by the areas populated by points.  These are
the galaxy positions for a narrow range in redshift from one of our mock
catalogues.  The 2dF consists of two strips.  The larger crosses the SGP
while the small one runs close to the NGP.
The solid curve marks the boundary of the SDSS survey, which is an
ellipse centred close to the NGP.  We do not include the SDSS southern strips.
The grid indicates the RA and dec. coordinates.
}
\label{fig:map}
\end{figure*}

The specifications of both the 2dF and Sloan surveys may be slightly modified
after evaluating the results from the current period of test observations.
The areas which we have adopted are shown in Fig.~\ref{fig:map}
and defined below. 

The main SDSS area is an elliptical region centred at 
${\rm R.A.}= 12^\hr 20^\min $ $\delta=32.8^\deg$,
close to the North Galactic Pole (NGP) and covering  $3.11$ steradians.  
The minor axis of the ellipse spans $110^\deg$ and runs along a line
of constant R.A.  The major axis spans $130^\deg$.
Our mock catalogues do not include the strips in the Southern Galactic Cap
that will also be part of the SDSS redshift survey; larger simulation 
volumes are needed to model simultaneously the Northern and Southern SDSS.

The main 2dF survey consists of two broad declination strips.  The larger
is approximately centred on the SGP and covers the declination range 
$-22.5^\deg>\delta>-37.5^\deg$.  This declination range breaks into three
contiguous, $5^\deg$ wide strips, each with slightly different ranges in R.A., 
which from north to south are $21^\hr 48^\min <{\rm R.A.}<3^\hr 24^\min $,
$21^\hr 39.5^\min <{\rm R.A.}<3^\hr 43.5^\min $ and 
$21^\hr 49^\min <{\rm R.A.}<3^\hr 29^\min $.  The smaller
strip in the northern galactic hemisphere covers $-7.5^\deg<\delta<2.5^\deg$
and $9^\hr 50^\min <{\rm R.A.}<14^\hr 50^\min $.  Together
they cover a solid angle of $0.51$ steradians.  There is considerable
overlap between the northern slice and the area covered by the SDSS.

\subsection{The radial selection function}
\label{sec:select}

The galaxies of the 2dF survey are selected from the APM
galaxy survey and will be complete to an extinction corrected apparent
magnitude of $\BJ < 19.45$.  The SDSS will have galaxies selected
from its own multi-band digital photometry.  The primary selection 
will be made in the Gunn-$r$ band, and it will include a surface brightness
threshold to ensure that an adequate fraction of the galaxy light
goes down a $3^{\prime \prime}$ fibre (see \cite{gw95} for details).
For simplicity, and because our goal
is merely to match the geometry and depth of the two surveys,
we make our selection for both catalogues in the $\BJ$ band.
For the SDSS we adopt a magnitude limit of $\BJ < 18.9$ so as to
approximately reproduce the SDSS target of 900,000 galaxies in the 
survey area.  A mock catalogue from a $(600\hmpc)^3$ \nbody simulation
that mimics the SDSS selection function
in greater detail will be presented elsewhere
(Gott et al., in preparation; see also \cite{gw95}).
In addition to its primary galaxy sample, 
the SDSS will target a set of $\sim 100,000$
luminous red elliptical galaxies, to create a deep, sparse sample
that is approximately volume-limited to $z \sim 0.4$.  Similarly, the 2dF
programme includes a deep extension to $R \sim 21$ which will contain
$\sim 10000$ galaxies. We do
not attempt to model these samples because their median depths are larger
than our simulation cubes.  

In order to compute the radial selection functions of the surveys,
we adopt a Schechter function description of the $\BJ$ band luminosity
function,
\begin{equation} 
\frac{d \phi(L)}{d L} \, dL = \phi_\star \, 
(L/L_\star)^{{\alpha_\star}} \, \exp(-L/L_\star) \, dL/L_\star , 
\end{equation}
with absolute magnitude $M_{\BJ}= M^\odot_{\BJ} -2.5 \log_{10}(L/L_\odot)$.  
We relate the apparent magnitude $B_J$ of a galaxy at redshift $z$
to the corresponding absolute magnitude $M_{\BJ}$ at redshift $z=0$ using
\begin{equation}
\BJ \negthinspace
 = \negthinspace
{\rm e} + {\rm k} + 5 \log_{10}(d_L/\hmpc) + 25 + 
(M_{\BJ} - 5 \log_{10} h) .
\end{equation}
Here $d_L$ is the luminosity distance to redshift $z$ in the appropriate
cosmological model.
The term ``k'' denotes the so called k-correction, which
arises from the Doppler shift to the wavelength of the galaxy's spectral
energy distribution when viewed in the observer's frame. 
The term ``e'' describes the effect of luminosity evolution in the galaxy
as a result of a combination of passive evolution of the stellar populations
and star formation.  This model therefore allows for luminosity evolution,
but not for any change in the shape of the galaxy luminosity function,
which might occur as a result of galaxy merging or luminosity dependent
evolution.

Even over the relatively limited range of apparent magnitudes covered by
the APM survey, the galaxy number counts are a significantly steeper
function of apparent magnitude than is predicted by non-evolving models
(\cite{maddox90b}). In contrast The K-band galaxy counts have 
shown no evidence for such a steep slope 
\cite{gard97}, but recently Phillips \& Turner (1998) have used
a compilation of survey data to argue that at the brightest magnitudes
the K-band slope is as steep as that seen in the B-band.
Unless we live in a very large
underdense region or there exists some as yet unidentified systematic error
in the bright galaxy counts, some form of rapid galaxy evolution is
necessary.  The counts can be reproduced by a model with strong luminosity
evolution such as can be accommodated in eqn.~(4.2), but at somewhat
fainter magnitudes than those covered by the SDSS and 2dF surveys such a
model predicts a tail of high redshift galaxies that is not seen in deep
spectroscopic galaxy samples (e.g. \cite{colless90}).  Thus, a more
complicated form of evolution is required, either one in which different
galaxies evolve at different rates or one in which galaxies merge so that
the number of galaxies is not conserved.  The new redshift surveys
themselves will give important information on evolution of the galaxy
luminosity function. However, for the purposes of quantifying large
structure this is not a problem provided that the selection function can be
accurately determined. We have therefore adopted a simple model that
produces a selection function with similar depth to that which we expect
the surveys to have.

\begin{figure}
\centering
\centerline{\epsfxsize= 8.5 truecm 
\epsfbox[ 0 60 574 750]{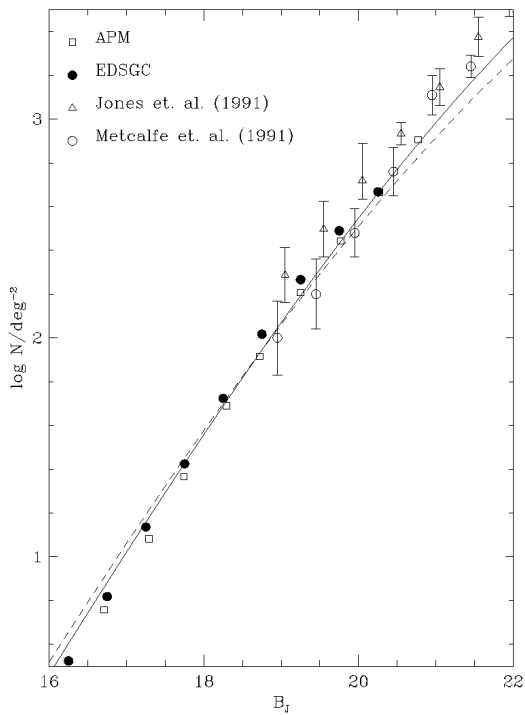}}
\caption{Galaxy number counts in our two evolution models compared with
observational data. Over this range of magnitudes, the counts are 
weakly dependent on cosmology and are plotted here for $\Omega_0=1$.
The solid line corresponds to our standard model
in which luminosity evolution cancels the k-corrections.  The dashed line 
corresponds to the less extreme model in which k-corrections are larger
than the luminosity evolution.  The data points are taken from
Maddox \etal (1990b) (APM),
Heydon-Dumbleton, Collins \& MacGillivray (1989) (EDSGC), 
Jones \etal (1991)
and Metcalfe \etal (1991).
}
\label{fig:counts}
\end{figure}

In our standard model we adopt the parameters found by Loveday \etal
(\shortcite{Loveday}) 
for the APM-Stromlo bright galaxy survey, $M^\star_{\BJ} - 5
\log_{10} h = -19.5$, $\alpha_\star=-0.97$ and $\phi_\star=1.4\times 10^{-2}
h^3 {\rm Mpc}^{-3}$.
We also set ${\rm k}+{\rm e}=0$, \ie we assume that strong luminosity evolution
occurs which cancels the k-correction. 
While this cancellation seems coincidental,
Fig.~\ref{fig:counts} shows that this simple choice gives reasonable
agreement with the observed galaxy number counts at $B_J \approx 19.5$
and so will produce mock galaxy catalogues with approximately the number
of galaxies expected in the 2dF survey.

\begin{figure}
\centering
\centerline{\epsfxsize= 8.5 truecm 
\epsfbox[0 0 570 530] {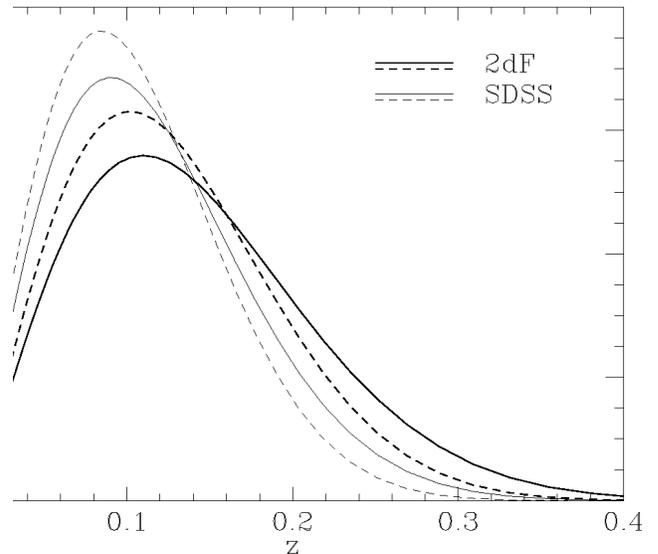}}
\caption{The model galaxy redshift distributions.  These distributions are 
weakly dependent on cosmology and are plotted here for $\Omega_0=1$.
The heavy curves, peaking at the higher redshifts, correspond
to the magnitude limit of $\BJ<19.45$ of the
2dF survey and light lines to the $\BJ<18.9$ of the SDSS.  
As in Fig.~\ref{fig:counts}, the solid curves are for our standard selection
function and the dashed curves for the alternative model with weaker luminosity
evolution.  The median redshifts are $z_{\rm m}=0.13$ and $0.12$
for the 2dF catalogues and $z_{\rm }=0.11$ and $0.10$ for the SDSS catalogues.
}
\label{fig:dndz}
\end{figure}

As a variation, we have also produced a selection of  mock catalogues 
in which the 
artificial assumption that ${\rm k}+{\rm e}=0$ has been dropped. 
For these we use the evolution law 
${\rm k}+{\rm e}= 2.5 \log_{10} (1+z)$,
which corresponds to weaker luminosity evolution than in our standard model. 
To compensate for this we increase the value of $\phi_\star$  by 24\% to 
$1.73\times 10^{-2 }h^3 {\rm Mpc}^{-3}$ to keep
the total number of galaxies in the
survey approximately the same as in our standard model.
This model's galaxy counts and redshift distributions (for the 
case of $\Omega_0 =1$)
are shown by the dashed lines in Figs.~\ref{fig:counts} and~\ref{fig:dndz}.

\subsection{Survey construction}

The task of generating a mock galaxy catalogue now consists of two steps:
choose the location of the observer, and select galaxies subject to 
the geometrical constraints and radial selection function specified above.

To aid in the comparison between the different cosmological models, we
choose to place the observer at the same position in each of the
galaxy catalogues.  The observer's position was essentially chosen at
random, although we did apply the weak constraint that the velocity
dispersion of particles within $5 \Mpc$ of the observer should be less
than $350 \kms$ in the $\Omega=1$ model, in order to avoid observers 
placed 
in rich galaxy clusters.  This constraint was only directly applied in
model E3S, but by virtue of the fact that all our simulations have the
same phases it is effectively satisfied in all the structure
normalized models. However, for the \COBE normalized simulations that
have $\sigma_8$ greater than that required to match the observed
abundance of rich clusters, the galaxy velocity dispersion is typically
higher, and the constraint may be violated. For most 
analyses of the 2dF and Sloan surveys the choice of the observer
should not be important, as the volumes of the surveys are large
compared to the local region whose properties are constrained by the
choice of observer.

Having chosen the observer's location, we replicate the periodic cube
of the \nbody simulation around the observer to reach a depth of
$z=0.5$. We choose the same position for the observer in both the
2dF and SDSS surveys, but the observer's orientation was not chosen
consistently between the two surveys.
We then loop over all the galaxies within the geometrical
boundaries of the survey.  From the model luminosity function and
cosmological model we compute the expected mean number density $\bar n_{\rm
s}(r)$ of galaxies brighter than the survey magnitude limit at the 
distance $r$ of each of these galaxies. We then
select the galaxy zero, one or more times according to a Poisson
distribution with mean $\bar n_{\rm s}(r)/\bar n_{\rm g}$, where $\bar
n_{\rm g}$ is the mean galaxy number density in the biased galaxy
distribution described in Section~\ref{sec:bias}.  
In this process approximately 1\% of the galaxies are selected more than once
and appear with identical positions and velocities in the mock catalogue. 
This double sampling essentially never occurs at $z>0.02$, where the
selection function drops to a space density less than $n_{\rm g}$.  For
each selected galaxy we generate an apparent $\BJ$ magnitude consistent
with the selection function, and also a value of $\zmax$, defined as the
redshift corresponding to the maximum distance at which the younger 
counterpart of the galaxy would still be brighter than the survey apparent 
magnitude limit.  In computing this redshift we include the effect of both 
the k-correction and evolution on the galaxy's luminosity.  As our idealized
models assume that galaxy mergers do not take place this definition
of $\zmax$ makes it easy to contruct volume limited catalogues in 
which the mean galaxy density is independent of redshift. 
For the genuine surveys removing the effect of evolution from the radial 
dependence of the galaxy density field
will be more problematic as evolutionary corrections for each galaxy
will be uncertain and over the limited redshift range probed by these surveys
galaxy mergers may also play a small role.
In our catalogues we record  the galaxy redshift, its
angular coordinates, the redshift it would have if it had no peculiar
velocity, its apparent $\BJ$ magnitude, and $\zmax$.  We also record an
index which can be used to identify the particle to which it corresponded 
in the original \nbody simulation.

\subsection{Adding long wavelength power}

For a subset of simulations we have applied a technique which allows
the spectrum of density fluctuations present in the final galaxy
catalogues to be extended to wavelengths longer than those included in
the original \nbody simulation.  This method, dubbed the Mode Adding
Procedure (MAP), was proposed by Tormen \& Bertschinger
(\shortcite{TB96}) and discussed further by Cole
(\shortcite{cole97}).  Essentially, one uses the Zel'dovich approximation
with a change of sign to remove from the \nbody particle distribution
the displacements caused by the longest wavelength modes in the
original simulation.  This can be done accurately if these modes are
still in the linear regime.  One then generates a new large scale
density field in a much larger box, which samples this same region of
$\k$-space more finely.   Displacements are computed by the Zel'dovich
approximation from this new field and used to perturb both the
original simulation cube and the adjacent replicas.  The displacements
applied to each of the replicas differ, as the new large scale
density field is not periodic on the scale of the original simulation
cube.  We choose to remove the inner $5^3$ modes from the original
simulations and generate the large scale density field in a box with
edge $7$ times that of the original simulation ($N_{\rm S}=2$ and
$L/S=7$ in the notation of Cole \shortcite{cole97}).

As pointed out by Cole (\shortcite{cole97}), it is important that the
biasing algorithm takes account of the effect of the added long wavelength
power.  This is most easily done for algorithms such as model~1, which are
a function of the initial linear density field.  One simply replaces the
original linear density field by a new one constructed by removing the
original long wavelength power and adding the new large scale density
field.  It is more complicated to correctly apply a biasing algorithm that
is a function of the final density field, because the final density field
is non-linear, and its short wavelength modes are coupled to the linear
long wavelength modes.  With this in mind, we applied the MAP only in
combination with our bias model~1.  In order to keep computer storage
requirements within reasonable bounds, it is necessary to combine into a
single program the application of the MAP, the biasing prescription, and
the survey selection criteria.

\subsection{Inventory}

For each of the cosmological simulations listed in Table~1 
(21, including the second realization of model E3S), 
we have created mock SDSS and 2dF surveys using bias model~1
and the standard selection function, in which the evolution and 
k-corrections cancel.  The MAP was not used to add long wavelength power
to these catalogues.  For four structure-normalized 
cosmological simulations --- the open $\Omega_0=0.4$
model (O4S), the flat $\Omega_0=0.3$ (L3S), and the two realizations of
the Einstein-de Sitter model (E3S) ---  we constructed a number of variants: 
changing the bias model to models~2, 3, 4, 5, and~6; without bias;
using the variation of the
selection function described in Section~\ref{sec:select},
in which luminosity evolution is weaker than the k-corrections;
and using bias model~1 with long wavelength power added using the MAP.

\section{Illustrations}
\label{sec:illust}

We now compare and contrast the visual properties 
of the galaxy distributions in each of the mock catalogues
using a series of redshift space wedge diagrams.

\begin{figure*}
\centering
\centerline{\epsfxsize= 17 truecm \epsfbox{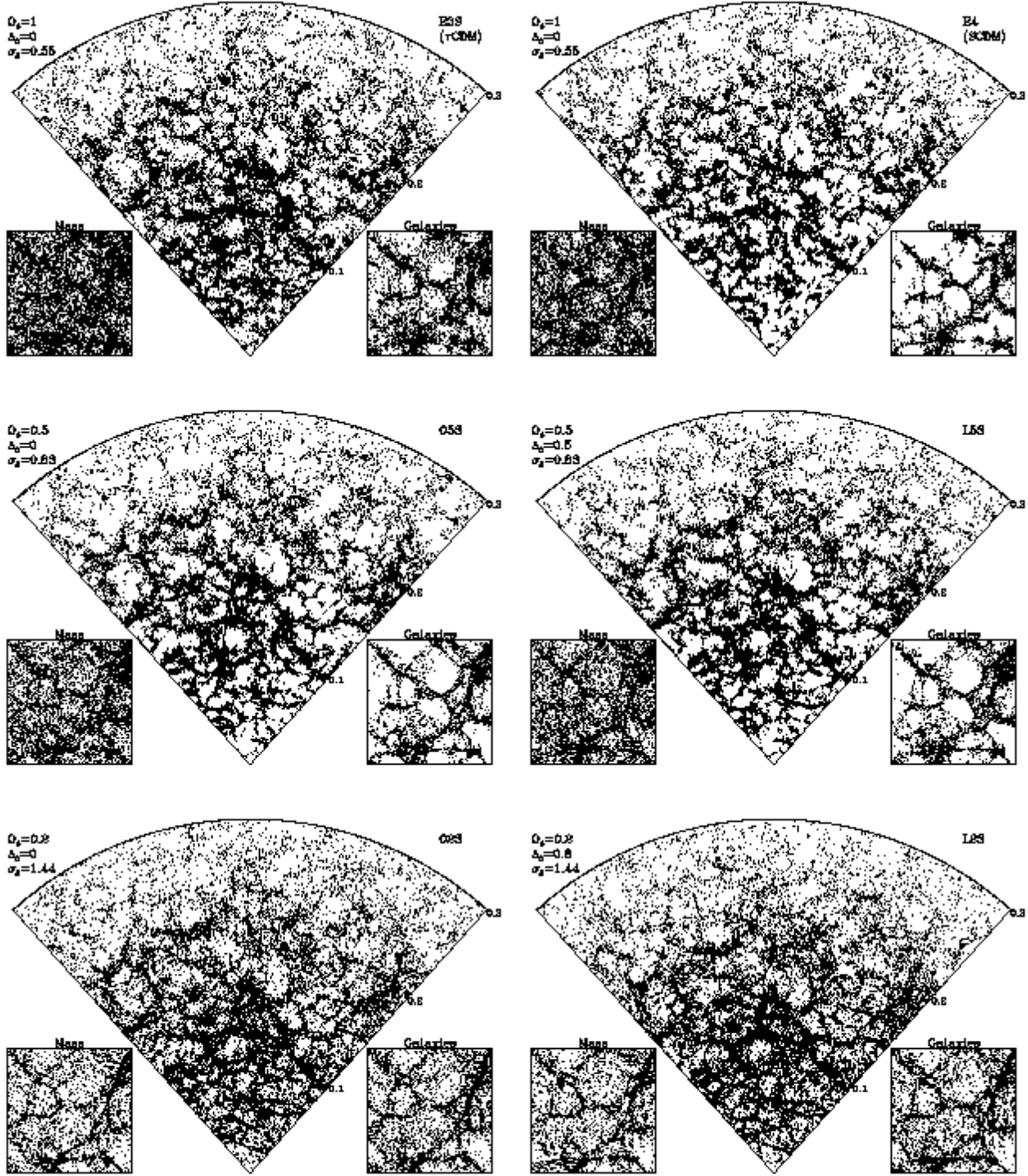}}
\caption{Redshift space slices showing galaxy positions from a variety
of the mock 2dF galaxy catalogues. Each wedge shows a strip $90^\deg$
wide in R.A. and $3^\deg$ thick in declination, extending to 
$z=0.3$.  Each of the six models shown is normalized by the present
abundance of galaxy clusters and biased using model~1
(see Section~\ref{sec:bias}). The inset square panels illustrate the effect of
bias by showing the real space particle and galaxy distributions in a
$100\times100\times20 \hmpc$ slab.  The top panels show
$\Omega_0=1$ models: E3S ($\tau$CDM) on the left and 
E4 (SCDM) on the right. Below these are the open and flat
$\Omega_0=0.5$ models, O5S and~L5S, and, at the bottom, the open and flat
$\Omega_0=0.2$ models, O2S and~L2S.  }
\label{fig:slice1}
\end{figure*}

\begin{figure*}
\centering
\centerline{\epsfxsize= 17 truecm \epsfbox{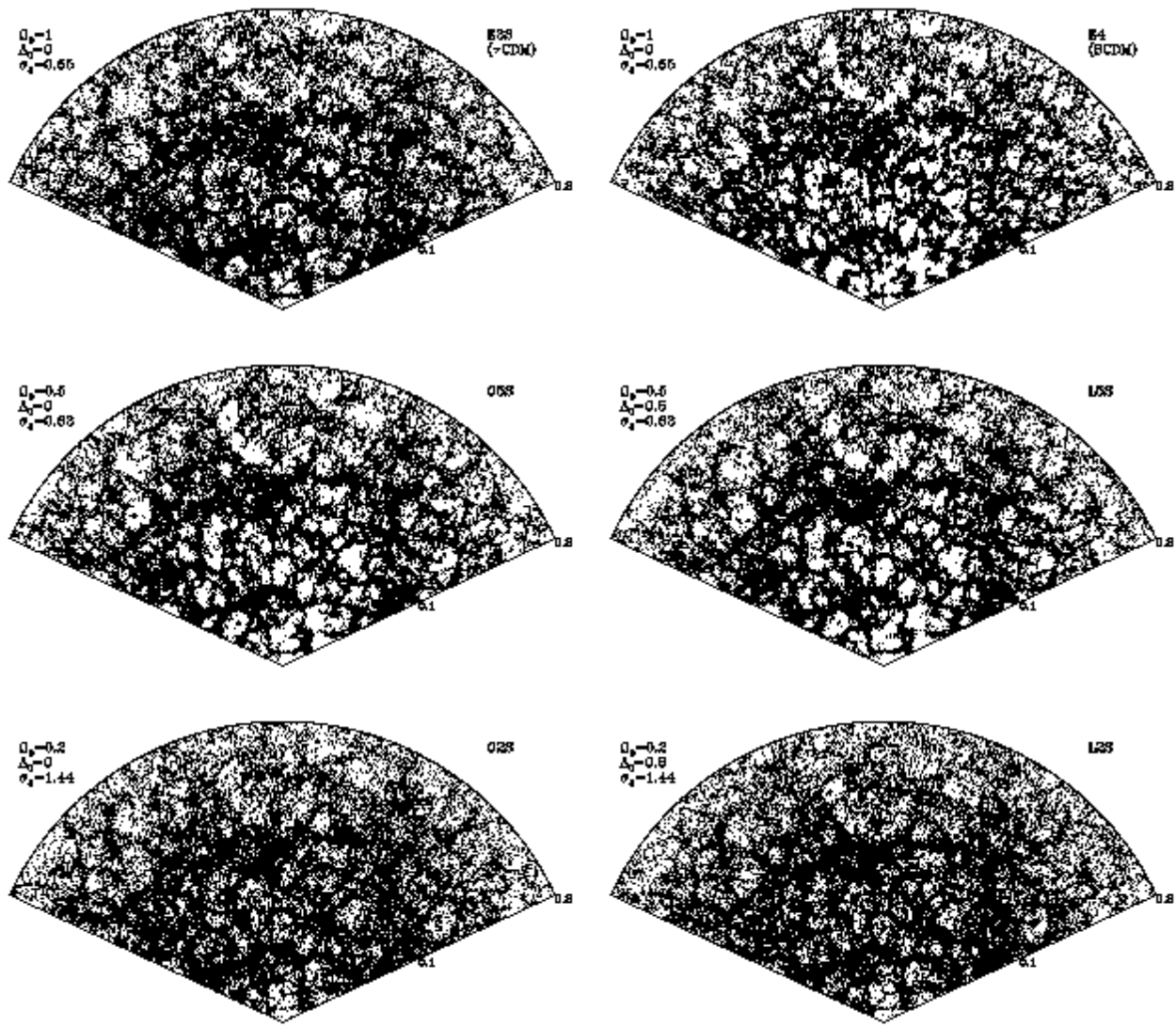}}
\caption{Redshift space slices from the cluster normalized mock SDSS
catalogues. The correspondence between model and panel is the same as
for Fig.~\ref{fig:slice1}. The slices are $130^\deg$ wide by $6^\deg$
thick and extend to $z=0.2$.  The qualitative
differences between the structure visible in these slices and in the
corresponding 2dF slices of Fig.~\ref{fig:slice1} are due to the choice
of slice thickness and depth rather than any intrinsic difference in
the 2dF and SDSS selection functions.  }
\label{fig:slice2}
\end{figure*}

\begin{figure*}
\centering
\centerline{\epsfxsize= 17 truecm \epsfbox{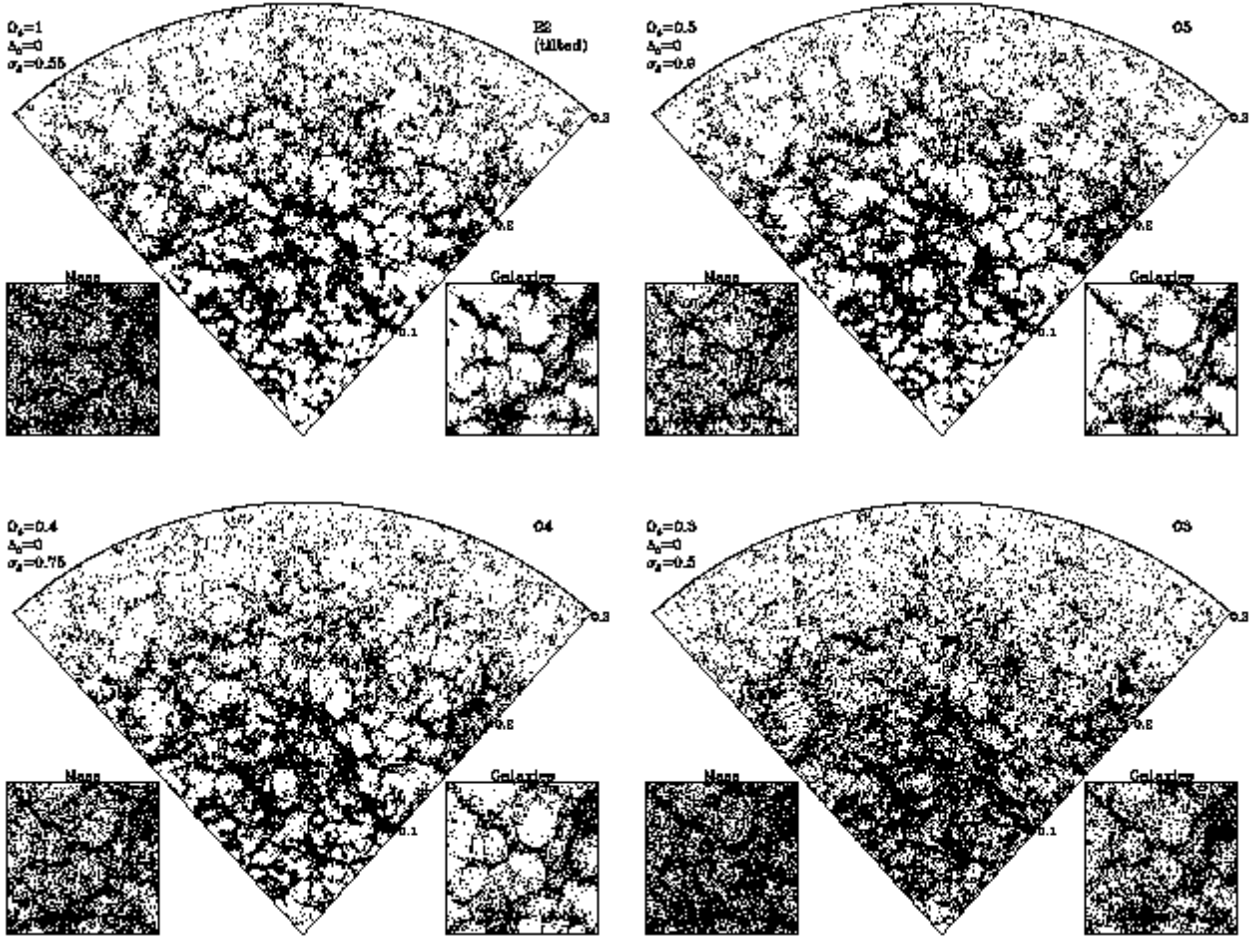}}
\caption{Redshift space slices from the mock 2dF catalogues for the
tilted CDM model E2 (top left) and the open \COBE normalized models,
O5, O4 and~O3. The corresponding value of $\Omega_0$ and the
normalization $\sigma_8$ are indicated on each panel.  The geometry of
the slices and inset plots of the real space mass and galaxy
distributions is the same as in Fig.~\ref{fig:slice1}.  }
\label{fig:slice3}
\end{figure*}

\begin{figure*}
\centering
\centerline{\epsfxsize= 17 truecm \epsfbox{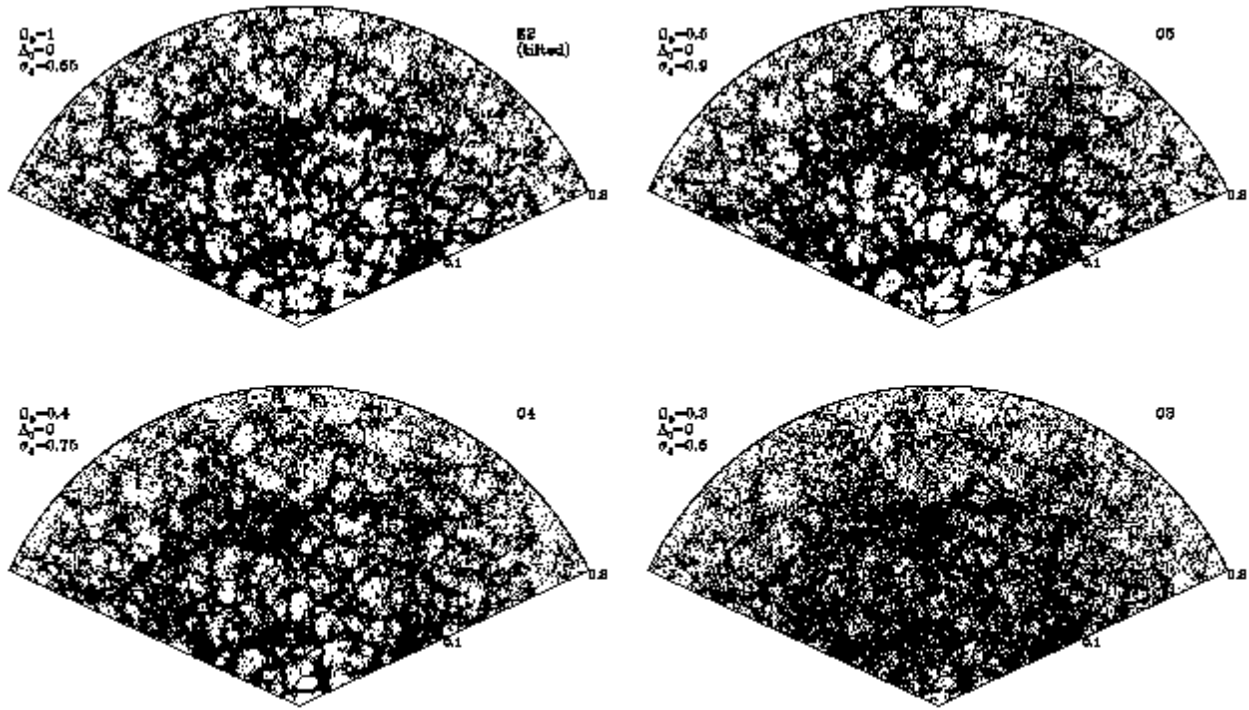}}
\caption{Redshift space slices from the mock SDSS catalogues for the
same models as Fig.~\ref{fig:slice3}, the tilted CDM model E2 and the
open \COBE normalized models, O5, O4 and~O3. The geometry of the slices
is the same as in Fig.~\ref{fig:slice2}.  }
\label{fig:slice4}
\end{figure*}

\begin{figure*}
\centering
\centerline{\epsfxsize= 17 truecm \epsfbox{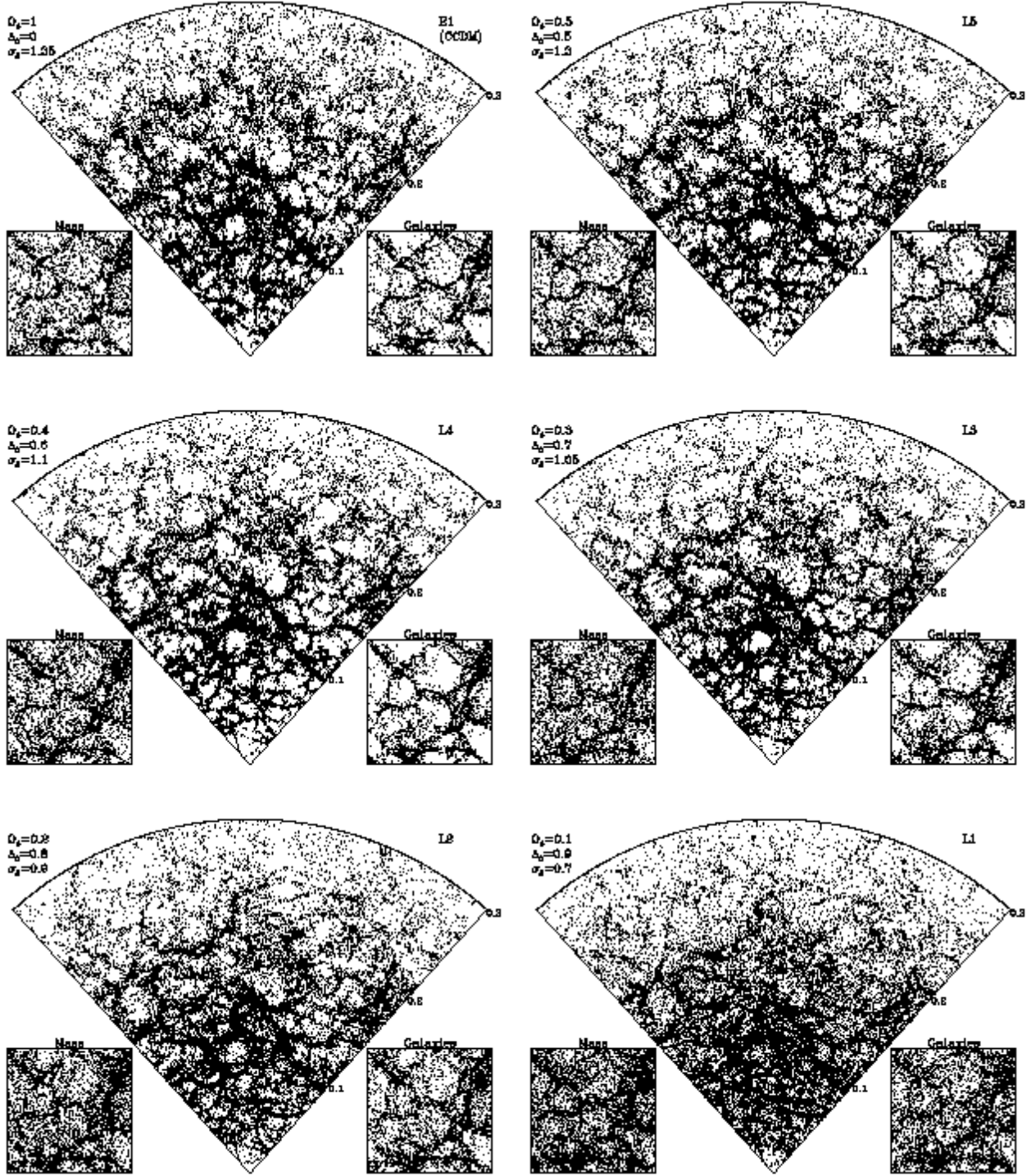}}
\caption{Redshift space slices from the mock 2dF catalogues for 
the flat \COBE normalized models, E1 (CCDM), L1, L2, L3, L4 and~L5.
The corresponding values of $\Omega_0$, $\Lambda_0$ and the
normalization $\sigma_8$ are indicated on each panel.  The geometry of
the slices and inset plots of the real space mass and galaxy
distributions are the same as in Fig.~\ref{fig:slice1}.  }
\label{fig:slice5}
\end{figure*}

\begin{figure*}
\centering
\centerline{\epsfxsize= 17 truecm \epsfbox{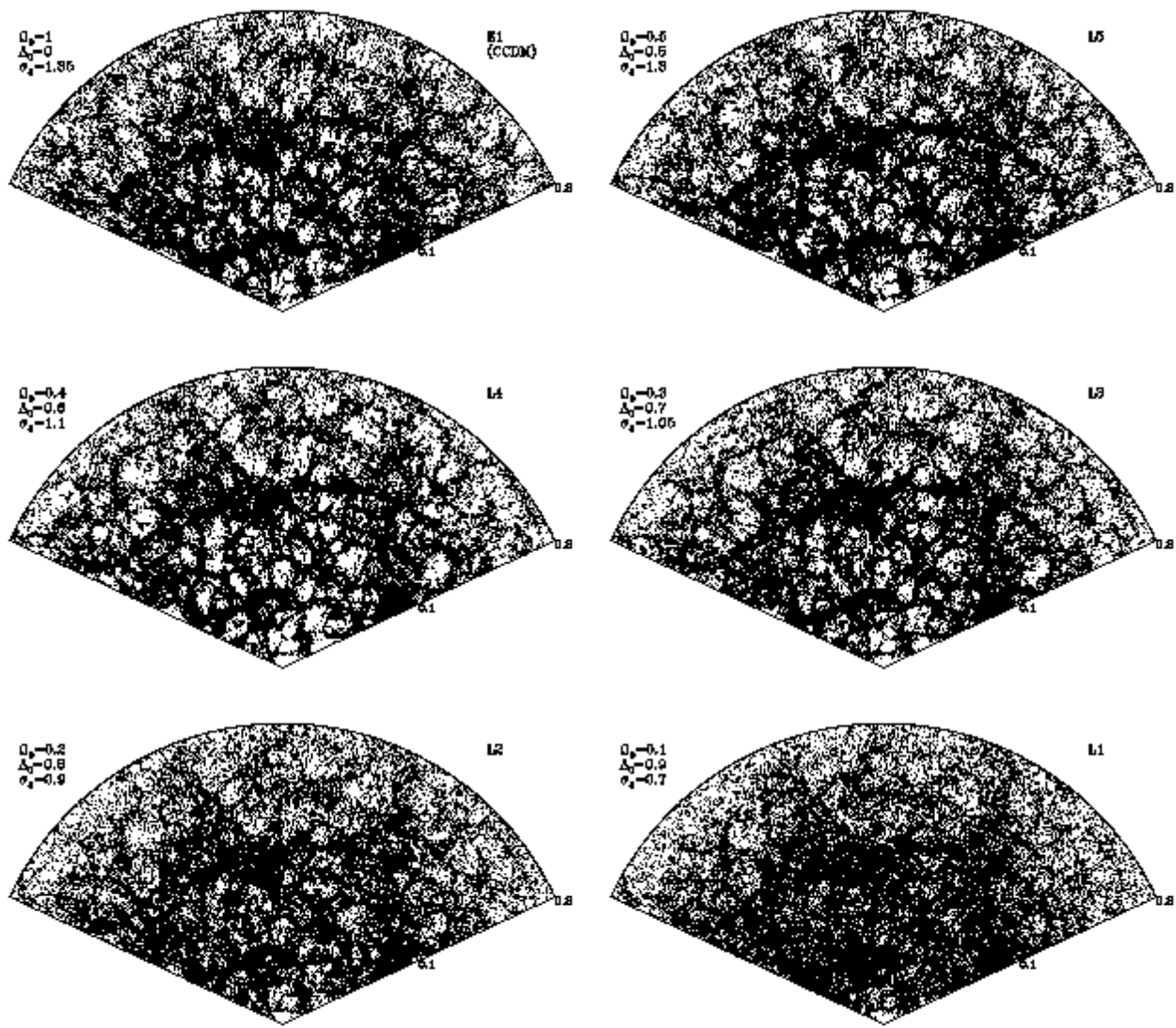}}
\caption{Redshift space slices from the mock SDSS catalogues for the
same models as Fig.~\ref{fig:slice5}, the flat
\COBE normalized models, E1 (CCDM), L1, L2, L3, L4 and~L5.
The geometry of the slices are the same as in Fig.~\ref{fig:slice2}.  }
\label{fig:slice6}
\end{figure*}

\begin{figure*}
\centering
\centerline{\epsfxsize= 17 truecm \epsfbox{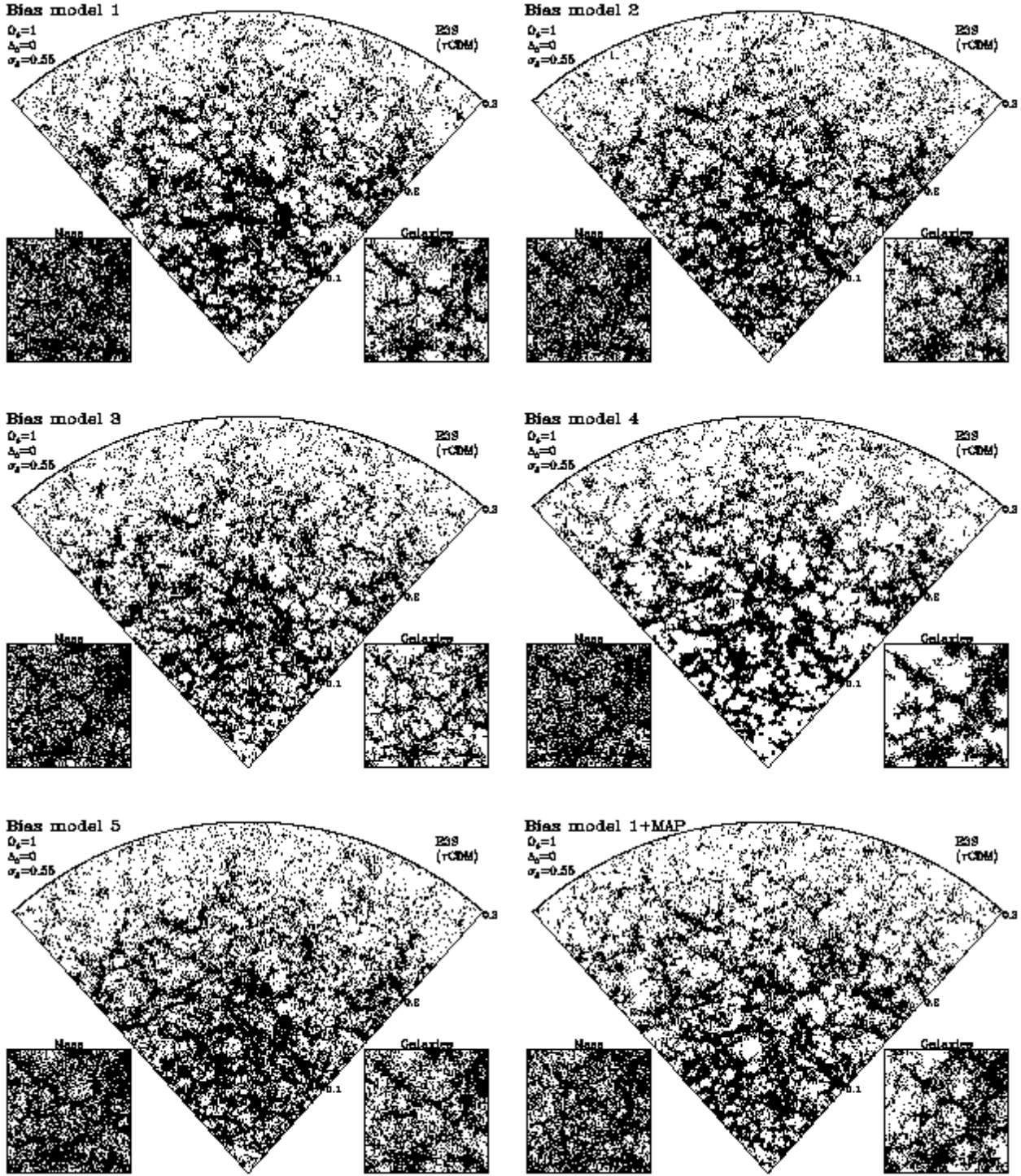}}
\caption{Redshift space slices from the mock 2dF catalogues showing
the effect of varying the choice of biasing algorithm. Each slice was
constructed from the same cosmological model E3S ($\tau$CDM), but with
a variety of biasing algorithms as indicated on each panel.
The panel at the bottom right shows the effect of using the MAP in conjunction
with bias model~1 to add long wavelength power to the mock
catalogue. The geometry of the slices and inset plots of the real space
mass and galaxy distributions are the same as in Fig.~\ref{fig:slice1}.  }
\label{fig:slice7}
\end{figure*}

\begin{figure*}
\centering
\centerline{\epsfxsize= 17 truecm \epsfbox{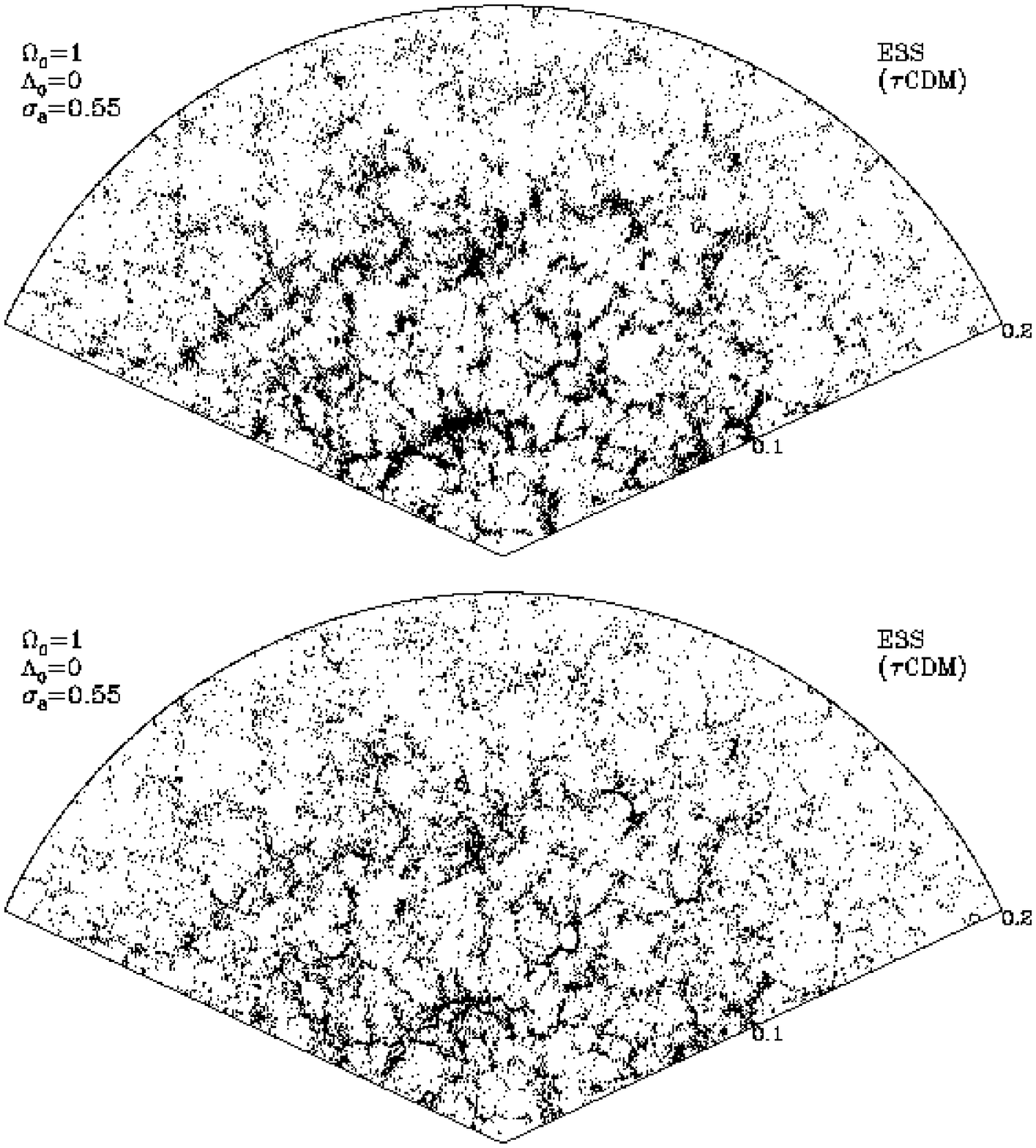}}
\caption{A comparison of the galaxy distribution in redshift space
(upper panel) and real space (lower panel) for a $2^\deg$ thick slice from
a SDSS mock catalogue constructed from model E3S ($\tau$CDM).}
\label{fig:slice8}
\end{figure*}

Figs.~\ref{fig:slice1} and~\ref{fig:slice2} show the galaxy
distribution in redshift space slices extracted from the 
mock 2dF and SDSS catalogues constructed from the cluster-normalized 
\nbody simulations. Each of the
catalogues was biased using model~1 of Section~\ref{sec:bias}.  The
2dF slices (Fig.~\ref{fig:slice1}) are $90^\deg$ wide in R.A., $3^\deg$
thick in declination and plotted out to a redshift of $z=0.3$. By 
contrast, the SDSS slices (Fig.~\ref{fig:slice2}), which are $130^\deg$
wide (corresponding to the full length of the long axis of the SDSS
ellipse) are $6^\deg$ thick but plotted only to $z=0.2$. A visual
inspection reveals that the structure in all six models looks
remarkably similar. This is essentially a reflection of the facts that all
the simulations were started with the same phases and that the observer
is alway located at the same position. Also, because these
models are designed to produce similar abundances of rich galaxy
clusters, the strength of the ``fingers-of-god'' effect is also similar.
The 1-dimensional galaxy velocity dispersions in all the cluster-normalized
models is in the range $440-465 \kms-$.
The visible effects on the galaxy distribution that result from varying
$\Omega_0$, $\Lambda_0$, and the amount of large scale power ($\Gamma$)
are quite subtle.  Of the two $\Omega_0=1$ models, E3S ($\tau$CDM) has
more large scale power than E4 (SCDM).  A manifestation of this is
that structure in E3S ($\tau$CDM) appears more connected and less
choppy than that of E4 (SCDM). The changes that occur when
$\Omega_0$ is varied are related to the strength of galaxy
biasing. For models that are normalized to produce the observed
abundance of rich clusters, the amplitude of mass fluctuations,
$\sigma_8$, increases as $\Omega_0$ is decreased. Thus, the $\Omega_0=1$
models require a strong bias, the $\Omega_0=0.5$ models a weak bias, and
the $\Omega_0=0.2$ models an anti-bias.  The effect of this can be
seen most clearly in the inset square panels of
Fig.~\ref{fig:slice1}. These show, in real space, a $100\times 100\times
20\hmpc$ slab of the mass and corresponding galaxy distribution, both
sampled to the same density of $\bar n_{\rm g} \approx 0.05 h^3{\rm
Mpc}^{-3}$. In the $\Omega_0\ge 0.5$ models, the biasing algorithm
clearly has the effect of mapping underdense regions in the mass
distribution to completely empty voids in the galaxy distribution.  In the
anti-biased, $\Omega_0=0.2$ models, galaxies continue to trace the mass
in the underdense regions.  Finally, comparison of the open and flat
models indicates that the value of the cosmological constant, $\Lambda_0$,
has virtually no detectable effect on the galaxy distribution.

Figs.~\ref{fig:slice3} and~\ref{fig:slice4} show redshift space slices with
the same geometry as those of Figs.~\ref{fig:slice1} and~\ref{fig:slice2}.
The top left hand panels in each figure show the tilted $\Omega_0=1$ model,
E2, which by virtue of the tilt is both cluster and \COBE normalized. These
distributions should be compared with those in the upper panels of
Figs.~\ref{fig:slice1} and~\ref{fig:slice2}, which show corresponding
slices for our other two cluster normalized, $\Omega_0=1$ models.  The
tilted (E2) model appears intermediate in character between the $\tau$CDM
(E3S) and SCDM (E4) models. This is consistent with the relative amounts of
power on scales of $50$-$100 \hmpc$ in these models. The tilt of 
$n\approx 0.8$ with $\Gamma \approx 0.45$ produces more
power on these scales than SCDM with $n=1$ and $\Gamma=0.5$, but less than
$\tau$CDM with $n=1$ and $\Gamma=0.25$. The remaining three panels in
Figs.~\ref{fig:slice3} and~\ref{fig:slice4} are for the open
($\Lambda_0=0$) \COBE normalized models. In this sequence, as $\Omega_0$ is
decreased $\sigma_8$ decreases, the bias increases, and $\Gamma$
decreases. The most visible effect comes from the variation of
$\sigma_8$. There is a clear trend such that the mass distribution looks
more evolved, with more crisply defined filaments and voids, as $\sigma_8$
is increased.  This trend is also visible in the galaxy distribution, but
here the bias partially compensates for the changing $\sigma_8$, and the
relationship appears weaker. On small scales the effect of the random 
velocities within galaxy clusters is just discernible. The 
``fingers-of-god'' are largest in the $\Omega_0=0.5$ in which the 
galaxies have a mean 1-dimensional velocity dispersion of $485 \kms-$
compared to only $225 \kms-$ in the $\Omega_0=0.3$

Figs.~\ref{fig:slice5} and~\ref{fig:slice6} show 2dF and SDSS redshift
space slices for the set of \COBE normalized, flat
($\Omega_0+\Lambda_0=1$) models.  For this sequence of models,
$\sigma_8$ decreases weakly as $\Omega_0$ decreases. Thus we see a
weaker version of the same trend we noted in the open \COBE normalized
models. The higher $\Omega_0$ models have a more evolved density
distribution with more sharply defined voids and filaments.
Also there is a similar trend in the galaxy velocity dispersion and
the resulting ``finger-of-god'' features. The 1-dimensional
velocity dispersion is $200 \kms-$ for $\Omega_0=0.1$ and climbs to
$665 \kms-$ for $\Omega_0=0.5$. The `fingers-of-god'' are extremely
pronounced in the $\Omega_0=1$ model which has a 
1-dimensional velocity dispersion of $890 \kms-$.

Fig.~\ref{fig:slice7} shows 2dF redshift space slices
illustrating the effect of varying the choice of biasing algorithm. 
Each slice was constructed from the same cosmological model E3S ($\tau$CDM), 
but with a variety of biasing algorithms as indicated on each panel. 
The correlation functions of each of these galaxy distributions,
shown in Fig.~3, are quite similar. Despite this some of the distributions
are visually quite distinct. The most striking feature is variation in the
size and number of voids. The voids are largest and most numerous in bias
model~4 as a result of its sharp density threshold. The models in which
the bias function is a more gradual function of density, such as the power 
law case of model~5, have far fewer voids.
The panel at the bottom right shows the effect of using the MAP in
conjunction with bias model~1 to add long wavelength power to the mock
catalogue. The distortion of the small scale galaxy distribution is small 
as the perturbations are of very long wavelength, but there effect on
measurements of large scale power can be appreciable.

Fig.~\ref{fig:slice8} contrasts the galaxy distribution in
redshift space (upper panel) with what would be observed if true 
distances rather than redshifts were measurable (lower panel). The model that
has been plotted here is the E3S ($\tau$CDM) model with galaxies
selected using bias model~1. The thickness of the slice is just $2^\deg$.

\section{Limitations}
\label{sec:limits}

We plan to use the mock catalogues presented in this paper to
help in the important task of testing and calibrating
the algorithms and statistics that will be applied to the analysis of the 
2dF and SDSS redshift surveys. We hope that they
will be similarly useful to other researchers.
However, it is important to be aware of the limitations 
of this compilation of mock catalogues.

\begin{enumerate}

\item The mock catalogues are idealized and do not suffer
    from some problems which, at some level, are inevitable in the
    genuine surveys. These include systematic errors in the photometry used
to select the target galaxies, cosmetic defects such as regions 
    cut out around bright foreground stars, failure to measure 
    redshifts for 100\% of the target galaxies, redshift measurement
    errors, and the residual effects of extinction by foreground dust. 

\item The model selection functions are simplistic
   and do not allow for the effects of galaxy mergers. It will only
   become possible to adequately constrain evolution models that incorporate 
   galaxy merging once the joint apparent magnitude-redshift distributions
   are accurately measured from the surveys themselves.
   Furthermore, we have not attempted to mimic the details of the 
   SDSS target selection criteria, although we expect that the
   selection function of the SDSS will not differ substantially from
   that implied by the $B_{\rm J}$-magnitude limited criterion that
   we have used.

\item Evolution of clustering  over the redshift range of the
     surveys is ignored --  each of our mock catalogues is constructed
     from a single output time from the \nbody simulations. Clustering
     evolution is probably very weak over the depth of the SDSS and 2dF
     surveys but it may not be negligible for deeper surveys and
     will be important for some applications (see, \eg
     \cite{nakamura98}).

\item The \nbody simulations solve the equations describing Newtonian 
    gravity and therefore explicitly ignore space curvature
    across the simulation box. One consequence of this 
    is that in the open models we are forced to use
    $4 \pi r_{\rm c}^2 dr_{\rm c}$, where $r_{c}$ is the comoving
    distance to redshift $z$, as the volume element rather
    than the correct relativistic expression. However for the depth of
    the present surveys this is a very small effect.

\item The simulations have limited mass and force resolution.
     The spatial resolution in the initial conditions is limited
     to scales greater than the mean particle separation of $1.8
\hmpc$. However, the power on these scales in the final configuration is
dominated by non-linear transfer from large scales. Thus, the range
of reliability of the estimated correlation functions and power
spectra is determined by the force resolution, $\epsilon = 90 \hkpc$ 
(comoving), and the particle mass, 
     $m_{\rm p}= 1.64\times 10^{12} \Omega_0 {\rm h}^{-1} {\rm M}_\odot$. 
The smallest structures that are resolved are galaxy groups and clusters. 

\item Because of the finite size of the \nbody simulation volume, $\k$-space
    is coarsely sampled and, in the absence of 
    the MAP extension, the catalogues have no power in wavelengths 
    $\lambda > 345.6 \hmpc$.  Since the depth of the surveys is 
    comparable to the size of the \nbody simulations, the coarse
    sampling could be problematic if one were to estimate the power spectrum
    from the mock catalogues using a high resolution estimator at
    values of $\k$ which do not match modes in the original
    simulation. There should be no problems for clustering statistics,
    such as the correlation function, which contain contributions from
    a broad range of $\k$.

\item The application of the MAP extends the power coverage in the mock 
   catalogues to wavelengths as large as $\lambda = 2420 \hmpc$ and improves 
   the sampling of $\k$-space at low $k$ ($k \lsim 0.026 {\rm h Mpc}^{-1}$), 
   but the sampling of $\k$-space remains coarse at larger $k$. Also,
   the MAP slightly modulates the frequencies of the existing high-$k$
   modes, with the result that although the high-$k$ power is
   still peaked around the modes present in the original simulation,
   some power is distributed to neighbouring values of $\k$. 
   Thus, narrow band estimates of the power spectrum at high $k$ may 
   still be slightly affected.

\item The mock catalogues assume galaxies trace the velocity
   field of the dark matter and thus that there is no velocity
   bias in the sense discussed, for example, by Carlberg, 
   Couchman \& Thomas (\shortcite{carlberg90}).

\item The adopted models of spatial bias are at best simplifications of
   the complex physics of galaxy formation.  Since reliable {\it a priori}
   predictions of bias are not possible with current simulation techniques,
   we have given each of our adopted cosmological models a ``good chance''
   by choosing bias parameters that force-fit the amplitude and
   (to the extent possible) the shape of the observed galaxy correlation
   function.  Our logic is that if the cosmological model in question is 
   to be consistent with current galaxy clustering data, then the
   ``true'' description of galaxy formation must somehow achieve the
   same thing that our biasing prescription does.  In selected cases
   we have produced multiple mock catalogues with a variety of biasing
   algorithms, so that the sensitivity of methods to the details
   of biasing can be investigated.

\end{enumerate}

\section{Instruction Manual}
\label{sec:manual}

Each mock catalogue can be downloaded from our WWW site \WWW .
Included in these pages is a detailed description of the catalogue 
file format. Each of the SDSS catalogue files
occupies 24~Mbytes. The smaller 2dF SGP and NGP catalogues occupy
5.4 and 2.7~Mbytes respectively. For each catalogue file
there is an associated selection function file that tabulates the
expected number of galaxies and the number density of galaxies as a function
of redshift for each model. 
We have also made available a number of fortran
subroutines. The first can be used to read the mock catalogue files.
A second reads one of the tabulated selection functions and can be
used to used to generate random galaxy positions consistent
with the survey radial selection function and geometric boundaries.

The main catalogue files list 7 properties for each catalogued galaxy,
$x$, $y$, $z$, $\zrest$, $\BJ$, $\zmax$ and~$i_{\rm ident}$.  The first
three of these are Cartesian redshift coordinates, \ie the galaxy
redshift is $\zgal=(x^2+y^2+z^2)^{1/2}$ and two angular coordinates
are defined by the relations $\sin \theta=z/\zgal$ and
$\tan \phi=y/x$.  For the 2dF catalogues these angles are simply the
declination $\delta=\theta$ and Right Ascension ${\rm R.A.}=\phi$.  In
the case of the SDSS they instead give a latitude, $\theta$, 
and longitude, $\phi$, relative to a pole at
the centre of the SDSS survey region and with respect to the major
axis of the SDSS ellipse.  The quantity $\zrest$ is the redshift the galaxy
would have if it had no peculiar motion and was just moving with the
uniform Hubble flow.  The redshift space coordinates can be converted
to real space coordinates by simply scaling each component by the
ratio $\zrest/\zgal$.  The galaxy's apparent magnitude is given by
$\BJ$.  The maximum redshift at which the galaxy would enter into the
catalogue taking account of the k-correction and luminosity evolution
is $\zmax$.  Thus selecting galaxies with both $\zrest < \zcut$ and 
$\zmax > \zcut$ will produce a volume limited catalogue to redshift $\zcut$.
Note that such volume limited catalogues will have a 
mean comoving number density of galaxies which is independent 
of redshift. This occurs in our idealized models
because we have assumed that galaxy merging can be ignored over the
limited redshift range probed by the surveys
and because we have included both the k-correction 
and evolutionary correction in our definition of $\zmax$.
The last property, $i_{\rm ident}$, is simply an index which relates the 
galaxy to a particle in the original \nbody simulation.

\section{Discussion}
\label{sec:conclude}

We have constructed, and made publically available, a set of mock galaxy
catalogues constructed from \nbody simulations having the
geometry and selection function appropriate to the forthcoming SDSS and
2dF redshift surveys. Our main intention has been to generate an extensive
and flexible suite of artificial datasets which may be used to develop,
test, and fine-tune statistical tools intended for the analysis of the real
surveys and, eventually, for testing the real data against theoretical
predictions. To this purpose we have generated mock surveys from simulations
with a range of cosmological parameters and made a variety of (biasing)
assumptions for extracting galaxies from the \nbody simulations.

Our mock catalogues are restricted to CDM cosmologies with Gaussian initial
fluctuations, but with a range of values for the cosmological parameters
$\Omega_0$, $\Lambda$, $H_0$, spectral shape parameter, $\Gamma$, etc.  It
will be interesting in future to extend this kind of work to other
cosmological models, particularly models that do not assume Gaussian
initial fluctuations. At present it remains somewhat unclear which
non-Gaussian models will be the most profitable to investigate. In our CDM
simulations, the fluctuation amplitude is set in two alternative ways: by
matching the amplitude of cosmic microwave background fluctuations as
measured by \COBE (and extrapolated to smaller scales according to standard
assumptions) or by matching the observed abundance of rich galaxy
clusters. One of our models (tilted $\Omega_0=1$) is deliberately
constructed so as to match both of these constraints while two others
(open $\Omega_0=0.4$ and flat $\Omega_0=0.3$) come
close to doing so on their own right. Although our suite of 20 models is
far from providing a well-sampled grid in this multidimensional parameter
space, it does include many of the cosmological models currently regarded
as acceptable.

We have implemented a variety of biasing prescriptions, all of which are
designed to reproduce approximately the known APM galaxy correlation function
over a limited range of scales.
The motivation for providing alternative biasing schemes is to
enable tests of the sensitivity to these assumptions of statistics which
attempt to infer properties of the mass from the measured properties of the
galaxies. In the absence of reliable theoretical predictions for the
formation sites of galaxies, we have taken the pragmatic approach of using
simple formulae, with one or two adjustable parameters, to characterise the
probability that a galaxy has formed in a region where the density field
has a given value. We have considered both Lagrangian and Eulerian schemes
in which the galaxies are identified in the initial and final density
fields respectively. We have restricted attention to ``local biasing"
models in which the probability depends solely on the value of the field
smoothed in the local neighbourhood of a point. An interesting extension 
would be to implement non-local biasing prescriptions such as the
cooperative galaxy formation model of Bower \etal (1993).

Over the range of scales adequately modelled by our \nbody simulations
($\sim 1-10 \Mpc$), our 2-parameter biased galaxy distributions match the
APM data remarkably well in almost all the cosmological models we have
considered, including those in which an antibias is required on small
scales. In some cases, a 1-parameter model suffices to obtain acceptable
results. In all cases the bias in the galaxy distribution is
scale-dependent even over the relatively narrow range of scales covered in
our simulations. As discussed by Jenkins
\etal (1998), scale-dependent biasing is a requirement of all viable CDM
models, 
and it is encouraging that simple heuristic models that depend only
on local density can achieve this, albeit over a limited range of
scales. When using our mock catalogues it is important to bear in mind that
while the locations of the galaxies are biased, the velocities are not --
our galaxies are assumed to share the velocity distribution 
of the associated dark matter.

A number of extensions of our work are possible. One that we have already
implemented but not discussed in this paper is the construction of mock
catalogues with the properties of other surveys, particularly surveys of
IRAS galaxies like the 1.2 Jy (Strauss \etal 1990) and the PSCZ surveys
(Saunders \etal 1995). Mock catalogues of the latter are already available
at the same web address as our 2dF and SDSS mock catalogues.  There are
several ways in which our catalogues could be improved to overcome at least
some of the limitations discussed in Section~6. For example, better \nbody
simulations are certainly possible with current technology. Larger
simulations would be particularly advantageous, since the size of those we
have used here is comparable to the depth of the real surveys. The
1-billion particle ``Hubble Volume" simulation of a 2 Gigaparsec CDM volume
currently being carried out by the Virgo consortium (Evrard \etal in 
preparation) will certainly be large enough, and we plan to extract mock
catalogues from it shortly. An interesting aspect of this simulation is
that data are output along a light cone and so the evolution of clustering
with lookback time can be incorporated into the mock catalogues.
Clustering evolution is expected to be negligible in the main 2dF and SDSS
surveys, but it will be important in the proposed faint extensions of these
surveys and to QSO surveys.

A further improvement would be to construct ensembles of mock catalogues
from independent simulations of each cosmological volume. These would help
quantify the cosmic variance expected in the real surveys. As we discussed
in Section~3, sampling effects are still appreciable on large scales even
with the huge volumes that will be surveyed with the 2dF and SDSS data.  In
fact, the fundamental mode in our simulations had a noticeable stochastic
downward fluctuation which can confuse the comparison with data on large
scales. Although this sort of effect can be quantified analytically to
some extent, simulations are useful in order to check for the effects of
biasing.  Finally, within a given \nbody simulation, there are already
better ways of identifying galaxies than the simple heuristic biasing
formulae that we have used. These new methods consist of grafting into an
\nbody simulation the galaxy formation rules of semi-analytic galaxy
formation models (e.g. Kauffmann, White \& Guiderdoni 1993; 
Cole \etal 1994). Examples of this approach already
exist (Kauffmann \etal 1997; Governato \etal 1998), but extensive mock
catalogues are still to be constructed using this technique. The combined
\nbody/semi-analytic approach offers the advantage of producing realistic
catalogues that include internal galaxy properties such as colours,
star-formation rates, morphological types, etc. Such information would be
particularly valuable to exploit the photometric data of the SDSS survey.

We are planning to implement several of the improvements just mentioned and
to update our web page as we progress. In the meantime we hope that the
gallery of mock catalogues already available will be of use to researchers
interested in the 2dF and SDSS surveys.

\section*{Acknowledgements}

SMC acknowledges the support of a PPARC Advanced Fellowship,
SJH a PPARC Studentship and CSF a PPARC Senior Fellowship.
DW acknowledges support from NASA Grant NAG5-3111
and NSF Grant AST-9616822. This work was partially supported by the
PPARC rolling grant for extragalactic astronomy and cosmology at 
Durham.



\begin{thebibliography}{}
\bibitem[\protect\citename{Bardeen \etal }1986]{BBKS} 
  Bardeen J.M., Bond J.R., Kaiser N., Szalay A.S., 1986, ApJ, 304, 15
\bibitem[\protect\citename{Baugh }1996]{baugh96}
   Baugh C.M., 1996, MNRAS, 280, 267
\bibitem[\protect\citename{Baugh \& Efstathiou }1993]{baugh93}
   Baugh, C. M.,  Efstathiou, G., 1993, MNRAS, 265, 145
\bibitem[\protect\citename{Baugh \& Efstathiou }1994]{BE94} 
  Baugh C.M., Efstathiou G., 1994, MNRAS, 270, 183
\bibitem[\protect\citename{Baugh \etal}1995]{baugh95} 
  Baugh C.M., Gazta\~naga E., Efstathiou G., 1995, MNRAS, 274, 1049
\bibitem[\protect\citename{Bond  \& Efstathiou }1991]{be91}
  Bond, J. R.,  Efstathiou, G., 1991, Phys. Lett. B, 265, 245
\bibitem[\protect\citename{Bower \etal}1993]{bow93}
Bower, R.G., Coles, P., Frenk, C.S. and White, S.D.M., 1993, ApJ, 405, 403. 
\bibitem[\protect\citename{Carlberg, Couchman \& Thomas }1990]{carlberg90} 
  Carlberg, R. G., Couchman, H. M. P., Thomas, P. A., 1990, ApJL, 352, L29
\bibitem[\protect\citename{Carlberg \etal }1997]{carlberg97} 
  Carlberg, R. G., Yee, H.K.C., Ellingson, E., 1997, ApJ, 478, 462 
\bibitem[\protect\citename{Cen \& Ostriker }1992]{cen92}
  Cen, R., Ostriker, J. P., 1992, ApJ, 399, L113
\bibitem[\protect\citename{Cen \& Ostriker }1993]{cen93}
  Cen, R., Ostriker, J. P., 1993, ApJ, 417, 415
\bibitem[\protect\citename{Chaboyer \etal }1996]{cdkk96} 
  Chaboyer B., Demarque P., Kernan P.J.,  Krauss L.M., 1996, Science, 271, 957
\bibitem[\protect\citename{Cole }1997]{cole97} 
  Cole, S.,  1997, MNRAS, 286, 38
\bibitem[\protect\citename{Cole \etal }1994]{cole4}
  Cole, S., Arag\'{o}n-Salamanca, A., Frenk, C.S., Navarro, J.F., Zepf,
         S.E., 1994, MNRAS, 271, 781
\bibitem[\protect\citename{CWFR}]{CWFR97} 
  Cole, S., Weinberg, D.H., Frenk, C.S., Ratra,B., 1997, MNRAS, 
  289, 37~CWFR
\bibitem[\protect\citename{Colless \etal }1990]{colless90} 
 Colless, M., Ellis, R., Taylor, K., Hook, R.N., 1990, MNRAS, 244, 408 
\bibitem[\protect\citename{Couchman }1991]{ap3m91} 
  Couchman H.M.P., 1991, ApJ, 368, 23
\bibitem[\protect\citename{Davis \etal }1985]{DEFW} 
 Davis M., Efstathiou G., Frenk C.S. \& White S.D.M., 1985, ApJ, 292, 371 
\bibitem[\protect\citename{Efstathiou, Sutherland \& Maddox }1990]{ESM90} 
  Efstathiou G., Sutherland, W. J. , \& Maddox, S. J., 1990, Nature, 344, 705
\bibitem[\protect\citename{Efstathiou, Bond \& White }1992]{EBW92} 
  Efstathiou G., Bond J.R., \& White S.D.M., 1992, MNRAS, 258, 1P
\bibitem[\protect\citename{Efstathiou \etal }1985]{edwf85} 
  Efstathiou G., Davis M., White S.D.M., Frenk C.S., 1985, ApJS, 57, 241
\bibitem[\protect\citename{Eke, Cole, \& Frenk }1996]{ecf96}
  Eke, V. R., Cole, S., Frenk, C. S., 1996, MNRAS, 282, 263
\bibitem[\protect\citename{Frenk \etal }1996]{fews96}
  Frenk, C.S., Evrard, A.E., White, S.D.M. and Summers, F. 1996, ApJ, 472, 460
\bibitem[\protect\citename{Gardner \etal }1997]{gard97}
 Gardner, J.P., Sharples, R.M., Frenk, C.S. \& Carrasco, B.E., 1997, 
 ApJ 480, L99 
\bibitem[\protect\citename{Governato \etal }1998]{gov98}
  Governato, F., Baugh, C., Cole, S., Frenk, C.S., Lacey, C., Quinn, T. \&
Stadel, J. 1998, Nature, 392, 359 
\bibitem[\protect\citename{Gunn \& Weinberg }1995]{gw95}
  Gunn, J. E., Weinberg, D. H., 1995, in Wide Field Spectroscopy
  and the Distant Universe, eds. S. Maddox \& A. Arag\'on-Salamanca,
  (Singapore: World Scientific), 3 (astro-ph/9412080)
\bibitem[\protect\citename{Hatton \& Cole }1998]{hc98}
  Hatton, S.J., Cole, S. 1998, MNRAS, 296, 10.(astro-ph/9707186)
\bibitem[\protect\citename{Heydon-Dumbleton \etal }1989]{heydon89}
  Heydon-Dumbleton N. H., Collins C. A., MacGillivray H. T., 1989, 
  MNRAS 238, 379 (EDSGC)
\bibitem[\protect\citename{Jenkins \etal }1997]{j96}
  Jenkins, A., Frenk, C.S., Pearce, F.R., Thomas, P.A., Hutchings, R. 
  Colberg, J.M., White, S.D.M., Couchman, H.M.P., Peacock, J.A. \& 
  Efstathiou, G., 1997, in ``Dark Matter 1996: Dark and Visible Matter in
  Galaxies and Cosmological Implications'', eds M. Persic and P. Salucci,
  PASP Conference Series, p 348. 
\bibitem[\protect\citename{Jenkins \etal }1997]{virgo97}
  Jenkins A., Frenk, C.S., Pearce, F.R., Thomas, P.A., Colberg, J., White,
S.D.M., Couchman, H.M.P., Peacock, J.A., Efstathiou, G., and Nelson,
A.H. 1998, ApJ, 499, 20. (astro-ph/9709010).
\bibitem[\protect\citename{Jones \etal }1991]{jones91}
  Jones L. R., Fong R., Shanks T., Ellis R. S., Peterson B. A., 1991, 
  MNRAS, 249, 481
\bibitem[\protect\citename{Katz, Hernquist, \& Weinberg }1992]{khw92}
  Katz, N., Hernquist, L., Weinberg, D. H., 1992, ApJ, 399, L109
\bibitem[\protect\citename{Kauffmann \etal }1993]{k93}
  Kauffmann, G., White, S.D.M., Guiderdoni, B., 1993, MNRAS, 264, 201
\bibitem[\protect\citename{Kauffmann \etal }1997]{k97}
  Kauffmann, G., Nusser, A., Steinmetz, M., 1997, MNRAS, 286, 795. 
\bibitem[\protect\citename{Loveday \etal }1992]{Loveday}
  Loveday, J. Peterson, B.A., Efstathiou, G., Maddox, S.J., 1992, ApJ, 90, 338
\bibitem[\protect\citename{Maddox, Efstathiou \& Sutherland }1990a]{maddox90}
  Maddox S.J., Efstathiou G., Sutherland W.J., Loveday J., 1990a,  MNRAS, 
  242, 43P
\bibitem[\protect\citename{Maddox \etal }1990b]{maddox90b}
  Maddox S.J., Sutherland W.J., Efstathiou G., 
  Loveday J., Peterson B. A., 1990b,  MNRAS,  247, 1P
\bibitem[\protect\citename{Maddox \etal }1990]{apmI}
   Maddox S.J., Efstathiou G., Sutherland, W.J., Loveday, J., 1990, 243, 692
\bibitem[\protect\citename{Maddox \etal }1996]{apmIII}
   Maddox S.J., Efstathiou G., Sutherland, W.J., 1996, MNRAS 283, 1227
\bibitem[\protect\citename{Mann, Peacock \& Heavens }1998]{MPE97}
  Mann R.G., Peacock J.A., Heavens A.F., 1998, MNRAS 293, 209. (astro-ph/9708031)
\bibitem[\protect\citename{Metcalfe \etal }1991]{metcalfe91}
  Metcalfe N,, Shanks T., Fong R., Jones, L. R., 1991, MNRAS, 249,498
\bibitem[\protect\citename{Nakamura, Matsubara, \& Suto }1998]{nakamura98}
  Nakamura, T. T., Matsubara, T.,  Suto, Y., 1998, ApJ, 494, 13.
  (astro-ph/9706034)
\bibitem[\protect\citename{Peacock \& Dodds }1994]{PD94} 
  Peacock J.A., Dodds S.J., 1994, MNRAS, 267, 1020
\bibitem[\protect\citename{Phillips \& Turner }1998]{PT98} 
  Phillips L.A., Turner E.L., 1998, ApJ, submitted. (astro-ph/9802352)
\bibitem[\protect\citename{Renzini \etal }1996]{renzini96} 
  Renzini A., et al., 
  1996, ApJ, 465, L23
\bibitem[\protect\citename{Salaris \etal }1997]{salaris96} 
  Salaris M., Degl'Innocenti S.,  Weiss A., 1997, ApJ, 484, 986
\bibitem[\protect\citename{Saunders \etal }1995]{saunders95}
 Saunders, W., Sutherland, W.J., Efstathiou, G., Tadros, H., Maddox,
  S.J., White, S.D.M., Oliver, S.J., Keeble, O., Rowan-Robinson, M. and
  Frenk, C.S., 1995 , {\it The Point Source Catalog Redshift Survey}, in {
  Wide Field Spectroscopy and the Distant Universe}, The 35th Herstmonceux
  Conference, World Scientific, 88. 
\bibitem[\protect\citename{Scherrer \& Weinberg }1998]{sw98}
  Scherrer, R. J., Weinberg, D. H. 1998, submitted to ApJ (astro-ph/9712192)
\bibitem[\protect\citename{Smoot \etal }1992]{smoot92}  
  Smoot G., et al., 1992, ApJ, 396, L1
\bibitem[\protect\citename{Sugiyama }1995]{sugiyama95}  
  Sugiyama N., 1995, ApJS, 100, 281
\bibitem[\protect\citename{Summers, Davis, \& Evrard }1995]{summers95}
  Summers, F. J., Davis, M., Evrard, A. E., 1995, ApJ, 454, 1
\bibitem[\protect\citename{Strauss \etal }1990]{str90}
 Strauss, M.A., Davis, M., Yahil, A., Huchra, J.P. 1990, ApJ, 361, 49.
\bibitem[\protect\citename{Tormen \& Bertschinger }1996]{TB96}
  Tormen G., Bertschinger E., 1996, ApJ, 472, 14
\bibitem[\protect\citename{Walker \etal }1991]{walker91}  
  Walker T.P., Steigman G., Schramm D.N., Olive K.A.,  Kang H.-S., 
  1991, ApJ, 376, 51
\bibitem[\protect\citename{White \etal }1987]{wfde87}  
  White S.D.M., Frenk C.S., Davis, M., Efstathiou G., 1987, ApJ, 313, 505 
\bibitem[\protect\citename{White }1994]{white94} 
  White S.D.M., 1994, Les Houches Lectures (astro-ph/9410043)
\bibitem[\protect\citename{White \etal }1993a]{wef93}  
  White S.D.M., Efstathiou G.,  Frenk C.S., 1993, MNRAS, 262, 1023 

\end{thebibliography}
\end{document}